\documentclass[longauth]{aa} 
\usepackage{graphicx}
\usepackage{rotating}
\usepackage{pdflscape}
\usepackage{subfigure}
\usepackage{txfonts}

\usepackage{natbib}
\usepackage{enumerate}

\bibpunct{(}{)}{;}{a}{}{,} 

\begin{document}

   \title{The Gaia-ESO Survey:  Insights on the inner-disc evolution from open clusters.\thanks{Based on observations collected with the FLAMES  spectrograph at VLT/UT2 telescope (Paranal Observatory, ESO, Chile), for the Gaia-ESO Large Public Survey (188.B-3002).}}

\author{L. Magrini\inst{1}, 
S. Randich\inst{1}, 
P. Donati\inst{2,3},
A. Bragaglia\inst{2},
V. Adibekyan\inst{4},
D. Romano\inst{2},
R. Smiljanic\inst{5},
S. Blanco-Cuaresma\inst{6},
G.  Tautvai\v{s}ien\.{e}\inst{7},
E. Friel\inst{8},  
J. Overbeek\inst{8},
H. Jacobson\inst{9},
T. Cantat-Gaudin\inst{10,11},
A. Vallenari\inst{11},
R. Sordo\inst{11}, 
E. Pancino\inst{2,12},
D. Geisler\inst{13}, 
I. San Roman\inst{14}, 
S. Villanova\inst{13},
A. Casey\inst{15},
A. Hourihane\inst{15},  
C.~C. Worley\inst{15},  
P. Francois\inst{16},
G. Gilmore\inst{15}, 
T. Bensby\inst{17},
E. Flaccomio\inst{18},             
A.~J. Korn\inst{19},
A. Recio-Blanco\inst{20},  
G. Carraro\inst{21},                
M.~T. Costado\inst{23}, 
E. Franciosini\inst{1},
U. Heiter\inst{24},
P. Jofr\'e\inst{15},
C. Lardo\inst{25},                  
P. de Laverny\inst{20},    
L. Monaco\inst{22},                 
L. Morbidelli\inst{1},
G. Sacco\inst{1},            
S.~G. Sousa\inst{26,27},
S. Zaggia\inst{11}
}

\institute{
INAF - Osservatorio Astrofisico di Arcetri, Largo E. Fermi, 5, I-50125 Firenze, Italy
\email{laura@arcetri.astro.it} \and 
INAF - Osservatorio Astronomico di Bologna, via Ranzani 1, 40127, Bologna, Italy \and
Dipartimento di Fisica e Astronomia, Universit\'a di Bologna, via Ranzani 1, I-40127 Bologna, Italy \and
  Instituto de Astrof\'isica e Ci\^encias do Espa\c{c}o, Universidade do Porto, CAUP, Rua das Estrelas, 4150-762 Porto, Portugal \and
   Department for Astrophysics, Nicolaus Copernicus Astronomical Center, ul. Rabia\'{n}ska 8, 87-100 Toru\'{n}, Poland \and
Observatoire de Gen\`eve, Universit\'e de Gen\`eve, CH-1290 Versoix, Switzerland \and
Institute of Theoretical Physics and Astronomy, Vilnius University, Gostauto 12, 01108 Vilnius, Lithuania \and
Astronomy Department, Indiana University, 727 East 3rd Street, Bloomington, IN 47405, USA \and
Department of Physics and Kavli Institute for Astrophysics and Space Research, 
Massachusetts Institute of Technology, 77 Massachusetts Avenue, Cambridge, MA 02139, USA \and
Dipartimento di Fisica e Astronomia, Universit\'a di Padova, vicolo Osservatorio 3, I-35122 Padova, Italy \and
INAF-Osservatorio Astronomico di Padova, vicolo Osservatorio 5, I-35122 Padova, Italy \and
ASI Science Data Center, via del Politecnico SNC, 00133, Roma, Italy\and
Departamento de Astronom\'{i}a, Casilla 160-C, Universidad de 
Concepci\'{o}n, Concepci\'{o}n, Chile \and
Centro de Estudios de F\'{i}sica del Cosmos de Arag\'{o}n (CEFCA), Plaza 
San Juan 1, E-44001 Teruel, Spain \and
Institute of Astronomy, University of Cambridge, Madingley Road, Cambridge CB3 0HA, United Kingdom \and
GEPI, Observatoire de Paris, CNRS, Universit\'e Paris Diderot, 5 Place Jules Janssen, 92190 Meudon, France\and
Dept. of Astronomy and Theoretical physics, Lund university, Box 43, SE-22100 Lund, Sweden \and
Laboratoire Lagrange (UMR7293), Universit\'e de Nice Sophia Antipolis, CNRS,Observatoire de la C\^ote d'Azur, CS 34229,F-06304 Nice cedex 4, France\and
INAF - Osservatorio Astronomico di Palermo, Piazza del Parlamento 1, 90134, Palermo, Italy\and
Department of Physics and Astronomy, Uppsala University, Box 516, SE-75120 Uppsala, Sweden\and
European Southern Observatory, Alonso de Cordova 3107 Vitacura, Santiago de Chile, Chile \and
Departamento de Ciencias F\'{i}sicas, Universidad Andr\'es Bello, Rep\'ublica 220, 837-0134 Santiago, Chile\\
Instituto de Astrof\'{i}sica de Andaluc\'{i}a-CSIC, Apdo. 3004, 18080 Granada, Spain\and
Department of Physics and Astronomy, Uppsala University, Box 516, SE-751 20 Uppsala, Sweden\and
Astrophysics Research Institute, Liverpool John Moores University, 146 Brownlow Hill, Liverpool L3 5RF, United Kingdom\and
Centro de Astrof\'isica, Universidade do Porto, Rua das Estrelas, 4150-762 Porto, Portugal \and
Departamento de F\'isica e Astronomia, Faculdade de Ci\^encias, Universidade do Porto, Rua do Campo Alegre, 4169-007 Porto, Portugal }

\date{Received ; accepted }

 
  \abstract
{The inner disc, linking the thin disc with the bulge, has been somehow neglected in the past because of intrinsic difficulties in its study, due, e.g., to 
crowding and high extinction. Open clusters located in the inner disc are among the best tracers of its chemistry at different ages and distances.   }
{We analyse the chemical patterns of four open clusters located within 7~kpc of the Galactic Centre and of field stars  to infer the properties of the inner disc
with the Gaia-ESO survey {\sc idr2/3} data release. 
}
{We derive the parameters of the newly observed cluster, Berkeley~81, finding an age of about 1~Gyr and a Galactocentric distance of $\sim$5.4~kpc. 
We  construct the chemical patterns of clusters and we compare them with those of  field stars
in the Solar neighbourhood and in the inner-disc samples. 
  }
{Comparing the three populations  we observe that inner-disc clusters and field stars are both, on average, enhanced in [O/Fe], [Mg/Fe] and [Si/Fe]. 
Using the {\sc idr2/3} results of M67, we estimate the non-local thermodynamic equilibrium (NLTE) effect on the abundances of Mg and Si in giant stars. 
After empirically correcting for NLTE effects, we note that NGC~6705 and Be~81 still have a high [$\alpha$/Fe]. 
}
{The location of the four  open clusters and of the field population   reveals
that the evolution of the metallicity [Fe/H] and of  [$\alpha$/Fe] can be explained within the framework of a simple chemical evolution model: both [Fe/H] and [$\alpha$/Fe]  
of Trumpler~20 and of NGC~4815 are in agreement with expectations from a simple chemical evolution model. 
On the other hand, NGC~6705, and at a lower level  Berkeley~81, have  higher  [$\alpha$/Fe] than expected for their ages, location in the disc, and 
metallicity. These differences might originate from local enrichment processes as explained in the inhomogeneous evolution framework. 
}
\authorrunning{Magrini, L. et al.}
\titlerunning{\sc The chemical composition of  the inner disc}

   \maketitle

\section{Introduction}

Open clusters are among the best tracers of the chemical evolution of the Galactic thin disc \citep[see, e.g.,][]{friel95,bragaglia06} from the very distant outskirts  \citep[e.g.][]{sestito08,yong12} to 
the inner regions, close to the Galactic bulge  \citep[see, e.g.,][]{magrini10,magrini14}. 
In particular, the clusters in the so-called inner disc, i.e. the part of the Galactic thin disc located at Galactocentric distances $\leq$8~kpc\footnote{In the whole paper the adopted  $R_{\rm GC,\odot}$ is 8~kpc \citep[see][]{mal13}},  
have been so far relatively little explored: indeed,
few clusters are known and spectroscopically observed \citep[e.g][]{carretta05,carretta07,sestito07,magrini10,carraro14a, carraro14b, carraro14c, magrini14}. 
This is due both to observational limits, i.e., high reddening and crowding towards the Galactic Centre, as well as to high mortality of clusters 
in regions where the density of stars is higher 
\citep[][]{freeman70,van02}. 
However, this region is of great importance because it constitutes a connection between the properties of the bulge and of the thin/thick disc. 

The Gaia-ESO Survey \citep[][]{gilmore12,RG13} is providing high-resolution spectroscopy with FLAMES@VLT and homogeneous stellar parameters and abundances for a large sample of Galactic open clusters. 
The sample includes clusters located throughout the disc, 
at different distances from the Sun, and with ages ranging 
 from their initial phases (with the exception
of the embedded ones), i.e., a few Myr after their formation, 
up to several Gyr (Randich et al., in prep.). 
The selection of targets to be observed within clusters is
as uniform as possible: briefly, Gaia-ESO observes unbiased samples
of cluster candidates with GIRAFFE, i.e. their selection is based only on their location of the colour-magnitude diagrams and no other information, as radial velocity, is used to ensure 
their membership, and more secure
cluster members with UVES (see Bragaglia et al., in preparation for
more details). 

Observations for several clusters were completed in the first 18 months 
of the Gaia-ESO Survey.  Analysis is now completed and the products
have been released to the {\em consortium} as Internal Data Release~2, 
hereafter {\sc idr2/3}.
The sample includes four very interesting intermediate-age/old clusters in the inner disc: NGC 6705, Trumpler 20, NGC 4815, and Berkeley 81.
The first three have been discussed in a series of previous paper using Gaia-ESO {\sc idr1} recommended values \citep[e.g.,][]{magrini14,tautvasiene15}. In the present paper, 
we not only consider the updated {\sc idr2/3} analysis, with important differences in the determination of the final parameters with respect to {\sc idr1} \citep[see][for details]{smi14}, but we add to the sample
Berkeley~81 (hereafter Be~81), which is the innermost cluster in the {\sc idr2/3} sample.
We aim at comparing its abundance pattern with both
field stars located in the Solar neighbourhood and toward the Galactic centre,  and with the other three clusters located in the inner disc. 

Be~81 is an intermediate-age cluster located at approximately 6~kpc from the Galactic centre (l=34.505, b=-2.068). 
It has been recently studied by \citet{donati14}, who, comparing observed and synthetic colour-magnitude diagrams, found an
age close to 1~Gyr, a large reddening $E(B-V)=$0.9, a distance modulus of 12.4, and a metallicity close to Solar. 
Previous literature data on Be~81 are quite scarce: 
BVI photometry of a portion of Be~81 was obtained  by \citet{sagar98}, 
from which they derived an age of $\sim$1~Gyr and a reddening $E(B-V)=$1. 
The only determination of the metallicity of this cluster is by \citet{warren09} through calcium triplet spectroscopy. 
They obtained a rather uncertain sub-solar metallicity ([Fe/H]=-0.11$\pm$0.15~dex), somewhat unusual for its internal location in the Galactic disc. 
This is likely related to the limiting magnitude of \citet{warren09} that did not allow them to reach stars as faint as the cluster giant branch and so they likely were not able to identify any of the cluster member stars.

The paper is structured as follows: in Section~2 we describe the analysis of Be~81 in {\sc idr2/3}. In Section~3 we derive its fundamental parameters, whereas in 
Section~4 we compare the chemical patterns of field and cluster populations.  In Section~5 we discuss the chemical patterns, and in particular the $\alpha$-enhancement, in the context of our current knowledge of the  disc's evolution. In Section 6 we give our summary and conclusions.

\section{Sample and analysis of Be~81} 
\begin{figure}
   \centering
  \includegraphics[width=0.45\textwidth]{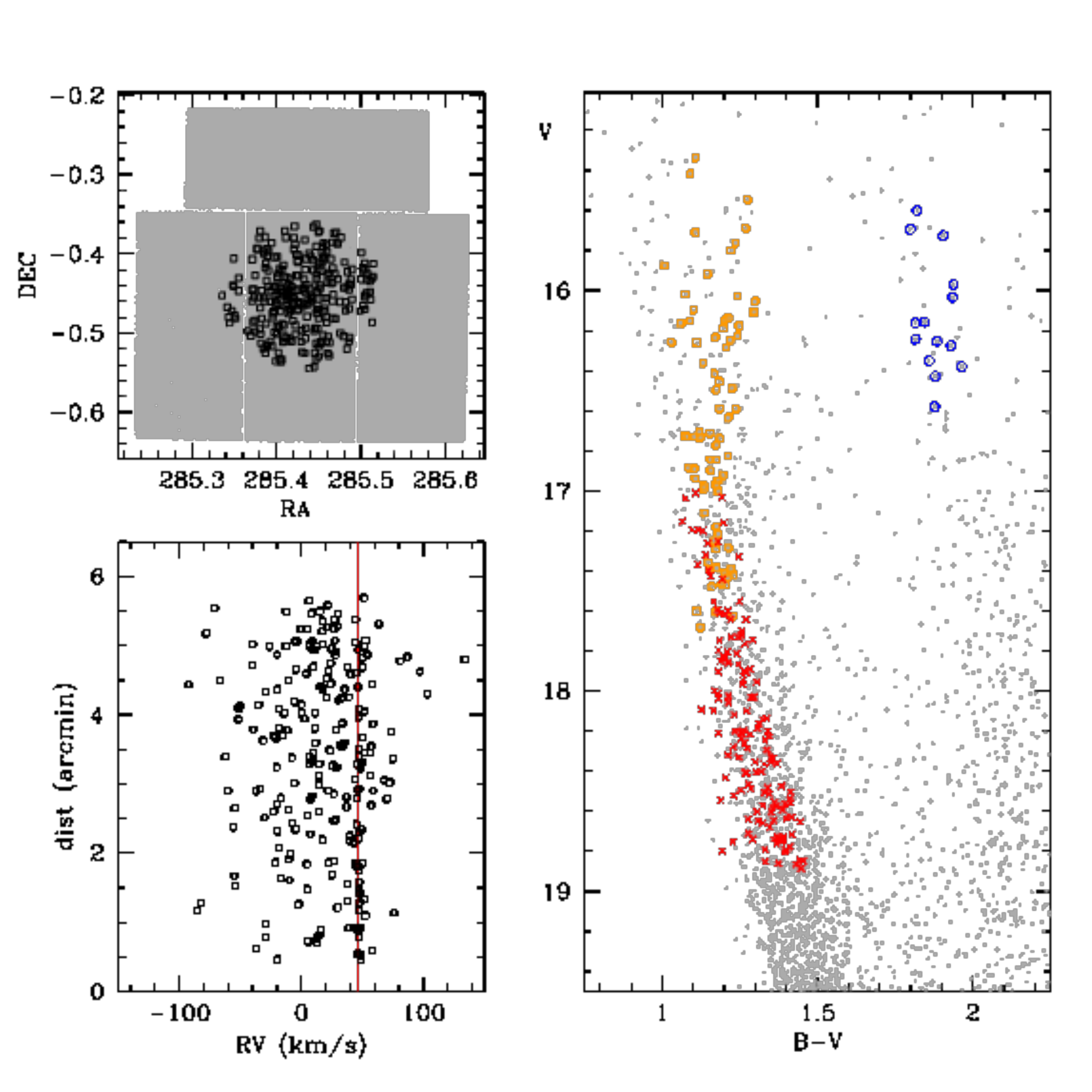}
    \caption{Upper left panel: in grey the Be~81 field of view of the photometric
data in Donati et al. (2014a), obtained with LBC@LBT; in black
the stars observed by the Gaia-ESO survey, all inside a radius of 6
arcmin. Lower left panel: distribution of the radial velocities versus
distance from the cluster centre. The red line is the cluster systemic
radial velocity. Right panel: CMD of the inner 6 arcmin, with stars
observed indicated by larger symbols (orange squares HR09B; red
crosses HR15N; blue circles UVES).
}
         \label{cmd}
   \end{figure}

In Be~81, 14 stars were observed with UVES using the setup U580 
(R=47000, $\lambda\lambda$~4800-5800 \AA); 266 stars were instead targeted
with  GIRAFFE using HR15N 
(R=17000, $\lambda\lambda$~6470-6790 \AA) and HR09B (R=25900, $\lambda\lambda$~5143-5356 \AA)  (16, 80, and 170 stars observed with both setups, HR09B only, 
and HR15N only, respectively).  
In Figure~\ref{cmd} (right panel) we show  the colour-magnitude diagram of the inner 6 arcmin of the cluster. The  UVES targets are {\em candidate} red clump giant stars, while the {\em candidate} main sequence stars were observed with  GIRAFFE.  In the upper left panel, the field of view of the photometric data in \citet{donati14} obtained with the Large Binocular Camera at the Large Binocular Telescope (LBC@LBT) is presented. 
Finally in the lower left panel, we show the  distribution of the radial velocities versus distance from the cluster centre. 

The spectrum analysis of the Gaia-ESO data is organised in several working groups (WGs) that provide astrophysical parameters and element abundances. 
In particular, the analysis of the  UVES and  GIRAFFE spectra discussed in the present paper was performed by three WGs: one dedicated 
to FGK-type stars observed with UVES (WG11-- see \citet{smi14}
for details), one for the same type of stars targeted with GIRAFFE 
(WG10-- Recio-Blanco et al.,~in prep.), and another one devoted to warmer
stars observed with both  UVES and  GIRAFFE setups (WG13--
Blomme et al.,~in prep.). 

The  atmospheric parameters obtained by  the
different WGs were then combined by a WG
dedicated to the homogenisation (WG~15) that provides a final set of recommended parameters (Hourihane et al.,~in prep). 
More specifically, the homogenisation process compares the results 
of the different WG for common targets and calibrators and, based on 
that, derives and applies offsets to obtain a common scale for all the Gaia-ESO Survey results. 
We mention that the {\sc idr2/3} WG~15 homogenisation process has not affected
the recommended values of effective temperature (T$_{\rm eff}$), surface gravity ($\log$~g) and  microturbolent velocity ($\xi$), while it has derived 
and applied offsets in metallicity [Fe/H] and in radial velocity (RV). 
Also, so far, the homogenisation flow of  {\sc idr2/3} does not involve the homogenisation of the elemental abundances through  different WG.  
Therefore, the elemental abundances used in the present work are those provided as recommended values by WG11 \citep{smi14}.

All the targets observed in Be~81 field have recommended atmospheric 
parameters and radial velocities in the final table of {\sc WG~15}. The UVES
targets also have element abundance determinations from WG11.
In Table~\ref{tab_par}  we report the information about the  
UVES stars: namely, we provide
their identification names ({\sc Cname})\footnote{the {\sc Cname} is an ID used in the Gaia-ESO Survey to univocally identify an object with its coordinates}  and {\sc id}\footnote{{\sc id} of the original photometric catalogue of \citet{donati14}, available also the CDS},  equatorial coordinates,  radial velocities,  stellar parameters with associated errors, 
RV from \citet{hayes14} with associated errors, and membership flag (described in the Section 2.2).
The {\sc idr2/3} RVs are in  good agreement with those of \citet{hayes14}. 
In  Table~\ref{tab_par_wg10}  we show the information about confirmed
members (based on their RV)
observed with  GIRAFFE and analysed by WG10 (see Section 2.2 for the membership probability definition). Their stellar parameters  are obtained with spectral indices \citep{damiani14} and thus  the microturbulent velocity $\xi$ is not determined. 
Stellar parameters of WG~13 are not used here
because of their large uncertainties, and therefore they are not presented in Table~\ref{tab_par_wg10}. 
\subsection{The field comparison sample}
\paragraph{The Solar neighbourhood} 
For the MW field stars, used as comparison in the present work, we considered the analysis of WG~11. 
Specifically,
the UVES Solar neighbourhood sample aims at obtaining an unbiased sample of $\sim$5000 G-stars within $\sim$2~kpc from the Sun in the turn-off phase (see Gilmore et al. in prep. for a description of the samples). The final
purpose is to quantify the local elemental abundance distribution functions in detail. For this paper we used the whole sample of 861  Milky Way solar neighbourhood stars with recommended parameters. 
\paragraph{The bulge and inner disc.}
UVES stars observed in parallel with the GIRAFFE targets dedicated to the study of the Galactic Bulge  are expected to be evolved stars belonging to both bulge and inner-disc populations. During the first 18 months of the Gaia-ESO Survey, 80 stars of the bulge/inner-disc were observed and are included in the present discussion.

\begin{table*}
\begin{center}
\caption{Parameters and their uncertanities for the UVES targets in Be~81. }
\scriptsize 
\begin{tabular}{crllllllllllllllll}
\hline\hline
{\sc Cname}     &   {\sc  ID} &R.A.       &     Dec.         &          RV   &$\delta_{\rm RV}$ &  T$_{\rm eff}$ &  $\delta_{\rm T_{\rm eff}}$ &$\log$ g &$\delta_{\rm log~g}$&[Fe/H] &  $\delta_{\rm [Fe/H]}$ &$\xi$  & $\delta_{\xi}$ &  RV$_{\rm HF14}$   &$\delta_{\rm RV_{\rm HF14}}$ & Mem. \\
		        &   			 &\multicolumn{2}{c}{J2000.0} &            (km/s)   &(km/s) &(K)&  (K) &&&& &&&(km/s) & (km/s)&  \\
\hline
   19013537-0028186 &      32696 &      285.397384 &      -0.471833 &      46.66  &       0.40    &       4966   &       88     &       3.09    &       0.17  &     0.3   &     0.1   &     1.49  &     0.12  &     -      & -      & m     \\
  19013631-0027447 &      47682 &      285.401294 &      -0.462442 &      46.94  &       0.40     &       4911   &       105    &       2.98    &       0.13  &     0.23  &     0.13  &     1.45  &     0.13  &     47.1    & 0.7     & m     \\
  19013651-0027021 &      33913 &      285.402107 &      -0.450607 &      48.0   &       0.40    &       5082   &       93     &       3.42    &       0.36  &     0.43  &     0.14  &     1.4   &     0.16  &     49.1    & 1.6     & m?    \\
  19013910-0027114 &      33770 &      285.412921 &      -0.453194 &      48.12  &       0.40    &       5015   &       42     &       3.04    &       0.12  &     0.15  &     0.04  &     1.48  &     0.09  &     48.2    & 0.8     & m     \\
  19013997-0028213 &      47704 &      285.416526 &      -0.472581 &      47.73  &       0.40    &       4880   &       73     &       2.66    &       0.2   &     0.23  &     0.22  &     1.48  &     0.07  &     47.5    & 1.2     & m     \\
  19014004-0028129 &      47688 &      285.416856 &      -0.470245 &      48.0   &       0.40   &       4963   &       78     &       3.08    &       0.41  &     0.25  &     0.1   &     1.45  &     0.19  &     -      & -      & m     \\
  19014127-0026444 &      46427 &      285.421947 &      -0.44569  &      46.28  &       0.40    &       4938   &       47     &       3.05    &       0.07  &     0.33  &     0.11  &     1.53  &     0.17  &     46.8    & 1.1     & m     \\
  19014194-0028172 &      47703 &      285.424747 &      -0.471459 &      47.52  &       0.40    &       4929   &       74     &       2.86    &       0.14  &     0.21  &     0.07  &     1.54  &     0.11  &     -      & -      & m     \\
  19014228-0027388 &      46358 &      285.426169 &      -0.460778 &      47.72  &       0.40    &       4907   &       57     &       2.8     &       0.15  &     0.22  &     0.08  &     1.53  &     0.12  &     47.2    & 0.7     & m     \\
  19014498-0027496 &      46477 &      285.437428 &      -0.463783 &      47.41  &       0.40    &       4769   &       70     &       2.82    &       0.18  &     0.22  &     0.08  &     1.47  &     0.12  &     47.8    & 0.7     & m     \\
  19014525-0023580 &      46431 &      285.438563 &      -0.399437 &      48.73  &       0.40    &       5022   &       12     &       3.04    &       0.09  &     0.2   &     0.08  &     1.48  &     0.09  &     48.1    & 1.3     & m     \\
  19014769-0025108 &      47674 &      285.448698 &      -0.419671 &      47.37  &       0.40    &       4827   &       58     &       2.81    &       0.26  &     0.14  &     0.06  &     1.48  &     0.19  &     47.2    & 0.6     & m     \\
  19015261-0025318 &      35423 &      285.469182 &      -0.425509 &      47.85  &       0.40    &       5015   &       83     &       3.14    &       0.18  &     0.29  &     0.14  &     1.46  &     0.11  &     48.2    & 1.0     & m     \\
  19015978-0028183 &      75596 &      285.499139 &      -0.471749 &      46.38  &       0.40    &       4771   &       38     &       2.44    &       0.08  &     -0.33 &     0.07  &     1.45  &     0.09  &     46.6    & 1.3     & nm    \\
\hline \hline
\end{tabular}
\label{tab_par}\\
\end{center}
\end{table*}

\begin{table*}
\begin{center}
\caption{Parameters  and their uncertainties for the GIRAFFE RV members Be~81 from WG~10.
 }
\scriptsize 
\begin{tabular}{crlllllllllll}
\hline\hline
{\sc Cname}     &   {\sc  Object} &R.A.       &     Dec.         &        RV   &$\delta_{\rm RV}$ &   SETUP& T$_{\rm eff}$ &  $\delta_{\rm T_{\rm eff}}$ &$\log$ g &$\delta_{\rm log~g}$&[Fe/H] &  $\delta_{\rm [Fe/H]}$ \\
		        &   			 &\multicolumn{2}{c}{J2000.0} &             (km/s)  & (km/s) &  & (K) & (K) &&  & & \\
\hline\hline
   19012975-0027170 &    42246 &    285.373896 &    -0.454711 &    48.79 &     1.39  &     HR15N &     6811  &     174   &     3.24  &     0.33  &     -0.38 &     0.18  \\
  19013178-0028535 &    32235 &    285.382405 &    -0.481533 &    45.63 &     0.41  &     HR15N &     7253  &     150   &     4.09  &     -     &     -     &     -     \\
  19013240-0023593 &    43014 &    285.385012 &    -0.399756 &    47.83 &     0.87  &     HR15N &     6458  &     150   &     4.28  &     -     &     -     &     -     \\
  19013287-0031312 &    41319 &    285.386942 &    -0.525359 &    44.74 &     0.51  &     HR15N &     6751  &     150   &     4.21  &     -     &     -     &     -     \\
  19013372-0025163 &    24010 &    285.390467 &    -0.421222 &    46.22 &     3.12  &     HR15N &     6628  &     142   &     3.97  &     0.28  &     0.00  &     0.12  \\
  19013495-0028320 &    32504 &    285.395616 &    -0.475558 &    45.51 &     0.72  &     HR15N &     6637  &     150   &     4.24  &     -     &     -     &     -     \\
  19013809-0027105 &    33782 &    285.408736 &    -0.452932 &    46.92 &     0.79  &     HR15N &     6751  &     150   &     4.21  &     -     &     -     &     -     \\
  19013886-0026309 &    34394 &    285.411909 &    -0.441952 &    48.29 &     3.29  &     HR15N &     6545  &     184   &     4.35  &     0.38  &     0.10  &     0.14  \\
  19014212-0025400 &    35245 &    285.425515 &    -0.427794 &    48.0  &     2.24  &     HR15N &     6840  &     126   &     4.36  &     0.24  &     0.32  &     0.11  \\
  19014417-0025115 &    45739 &    285.434035 &    -0.419915 &    47.76 &     2.2   &     HR15N &     6618  &     142   &     4.87  &     0.31  &     0.23  &     0.11  \\
  19014443-0026202 &    42487 &    285.43514  &    -0.438959 &    46.77 &     0.81  &     HR15N &     6562  &     91    &     4.23  &     0.21  &     0.66  &     0.07  \\
  19014532-0026401 &    44710 &    285.438867 &    -0.444473 &    47.02 &     0.49  &     HR15N &     6637  &     150   &     4.24  &     -     &     -     &     -     \\
  19014637-0024287 &    24366 &    285.443198 &    -0.407997 &    45.48 &     0.61  &     HR15N &     6751  &     150   &     4.21  &     -     &     -     &     -     \\
  19014728-0025559 &    35009 &    285.447001 &    -0.432221 &    42.74 &     2.32  &     HR15N &     6605  &     167   &     5.44  &     0.40  &     -0.17 &     0.15  \\
  19014772-0027550 &    22999 &    285.448838 &    -0.465302 &    46.76 &     1.14  &     HR15N &     6011  &     212   &     4.65  &     0.56  &     -0.44 &     0.18  \\
  19014982-0028078 &    22921 &    285.45759  &    -0.468819 &    43.66 &     3.45  &     HR15N &     6528  &     160   &     4.7   &     0.35  &     0.16  &     0.12  \\
  19015195-0023078 &    45059 &    285.466451 &    -0.385517 &    44.37 &     0.92  &     HR15N &     6676  &     90    &     4.63  &     0.2   &     0.60  &     0.08  \\
  19015300-0022239 &    25196 &    285.470846 &    -0.373304 &    44.58 &     0.88  &     HR15N &     6947  &     150   &     4.17  &     -     &     -     &     -     \\
\hline \hline
\end{tabular}
\label{tab_par_wg10}\\
\end{center}
\end{table*}

\subsection{Membership}

The determination of the RVs for GIRAFFE and UVES spectra, respectively, can be found in Koposov et al. (in preparation) and \citep{sacco14}.
We have used the recommended radial velocity of the  UVES and  GIRAFFE sample to define the cluster systemic velocity. 

\begin{figure}
   \centering
  \includegraphics[width=0.45\textwidth]{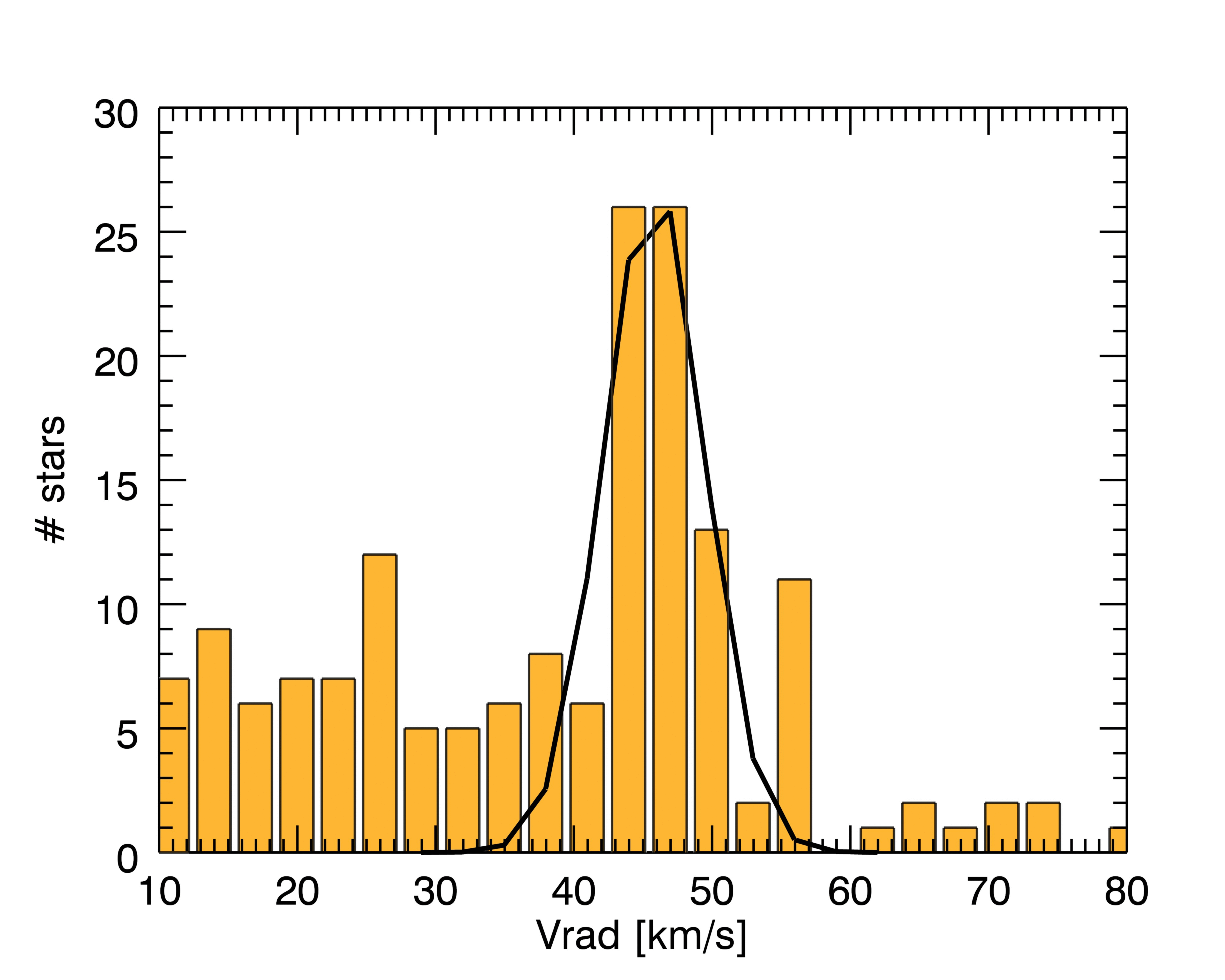}
    \caption{Distribution of radial velocities of the stars observed in the field of Be 81. The solid line is a gaussian fit of the distribution of the member stars. 
          }
         \label{Fig:histo:rv}
   \end{figure}

The resulting RV distribution, including both  GIRAFFE and  UVES 
targets, is shown in Fig.~\ref{Fig:histo:rv}. A gaussian fit of the peak of the distribution,  
identifying the cluster signature with respect to the field stars, gives a central velocity of  45.8 km~s$^{-1}$ and a Full Width Half Maximum (FWHM) of  3.1 km~s$^{-1}$. 
To avoid a large contamination by  field stars, we define a star as RV
cluster member if it has a RV within 1-$\sigma$ from the average cluster radial velocity. By doing so we identify
56 RV members, 14 observed with UVES and 42 targeted with GIRAFFE
(5 observed with both HR09B and HR15N, 11 with HR09B, 9 with HR15N and analysed by WG13, and 18 others observed with HR15N and analysed 
by WG10). In summary,
20\% of the observed targets (almost 100\% for  UVES and 16\% for GIRAFFE) have a RV consistent with membership.  
For  UVES, the contamination by field stars has been alleviated by observing high probability member stars whose radial velocity were obtained by \citet{hayes14}
to assist Gaia-ESO observations. 
For GIRAFFE, the low fraction of recovered cluster members is due
to the location of the cluster towards the Galactic Center where
 a large contamination by field stars is expected. 

\begin{figure*}
   \centering
  \includegraphics[width=0.95\textwidth]{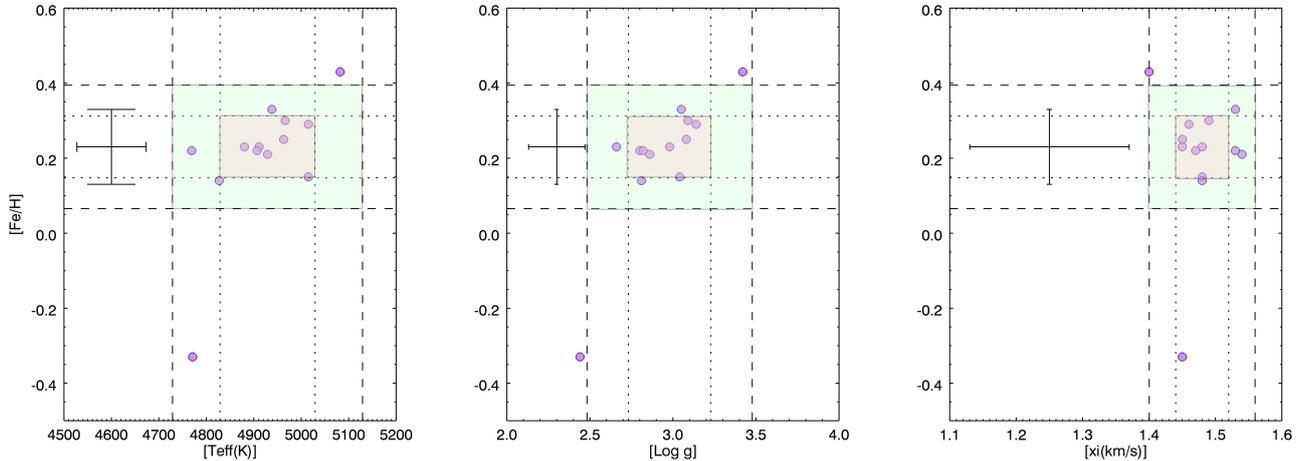}
    \caption{[Fe/H] vs. atmospheric parameters of  UVES stars in Be~81. The dotted lines mark the 1-$\sigma$ region, while the dashed lines indicate the 2-$\sigma$ area. 
    The typical error bar on each star are shown on the left side of each panel. }
         \label{Fig:par}
   \end{figure*}

In Fig.~\ref{Fig:par} we show [Fe/H] versus atmospheric parameters of the  UVES stars. We remind the reader that stars observed with  UVES are selected to be red giant stars in 
the clump (RC). 
They were, thus, expected to have similar atmospheric parameters. 
The average error (as given in the recommended value table) is plotted on the left side of each panel. 
The 1-$\sigma$ and 2-$\sigma$ around the median value \footnote{$\sigma$ is the median absolute deviation (MAD),  computed with the Robust sigma routine of IDL} areas are also marked. 
We note that the star with the lowest metallicity (19015978-0028183) is beyond 1-$\sigma$ from the median 
T$_{\rm eff}$ and 2-$\sigma$ from the median $\log$~g, likely being a non-member with 
radial velocity consistent with the cluster systemic 
velocity. 
This is not surprising  given 
the expected high contamination of field stars. 
The most metal rich star (19013651-0027021) is also somewhat deviant and might be a non-member star as well. 
In the next sections we discuss the chemical homogeneity of cluster member stars  and how it can help to disentangle 
member versus non-member stars.   

\begin{figure}
   \centering
  \includegraphics[width=0.45\textwidth]{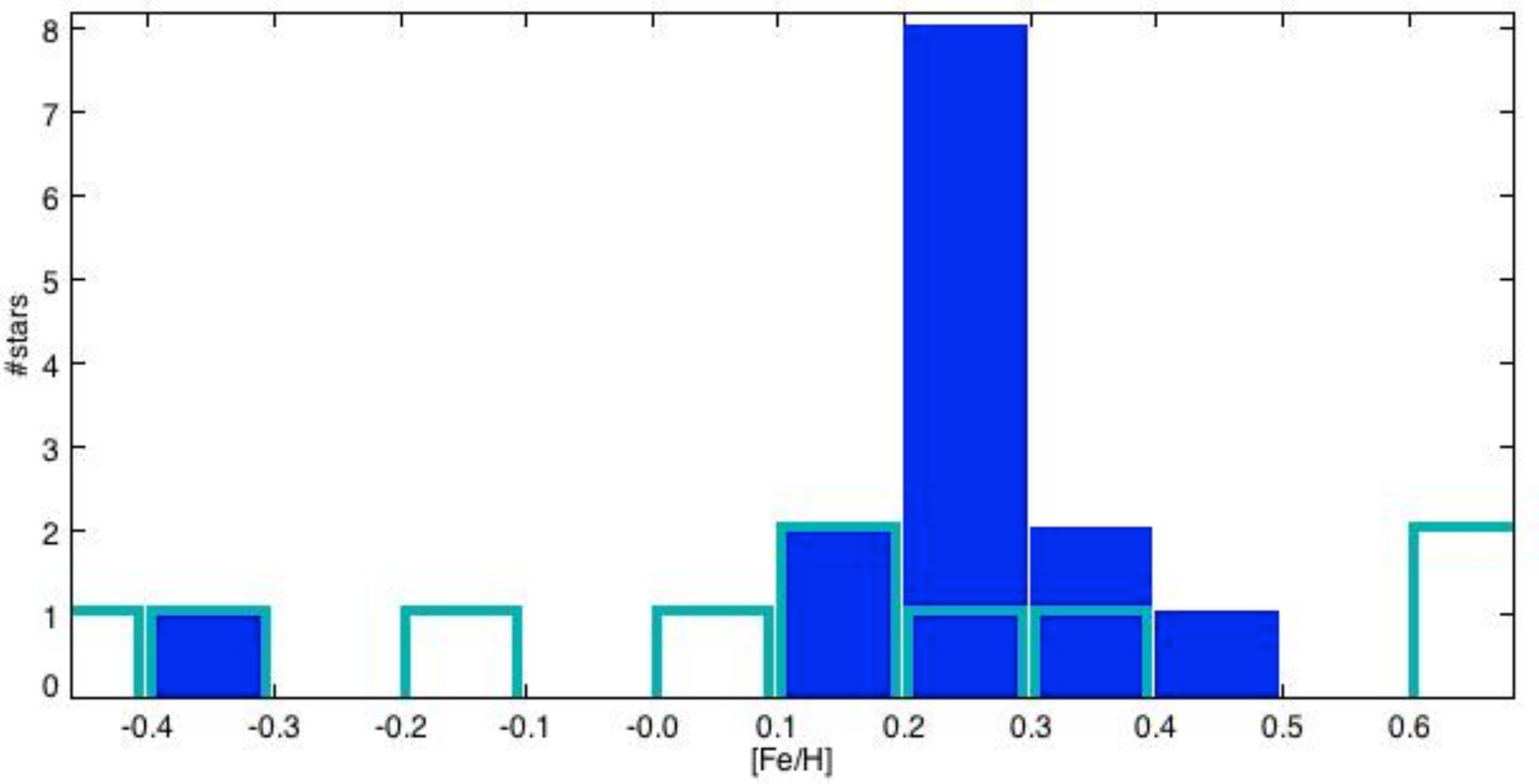}
    \caption{Distribution of [Fe/H] of the stars observed in the field of Be~81. The cyan rectangles show the  GIRAFFE metallicities of RV member stars, while the blue ones indicate  the  UVES metallicities (including  star 19015978-0028183).}
         \label{Fig:histo:feh}
   \end{figure}

In Fig.~\ref{Fig:histo:feh} we plot the histogram of the Gaia-ESO 
recommended  metallicities, [Fe/H], of the RV member stars observed with  UVES and  GIRAFFE. The two distributions are compatible: the median of the  UVES stars is [Fe/H]=0.23$\pm$0.08~dex, whereas for  the GIRAFFE stars is 0.16$\pm$0.44~dex. The UVES metallicities are 
much more concentrated 
in the bin centred at 0.25 dex, while the GIRAFFE metallicities are 
more dispersed due to the lower resolution and smaller spectral range. Note that the errors in Table~\ref{tab_par_wg10} are based on a single determination of stellar parameters, while the errors in Table~\ref{tab_par} are computed with the MAD of typically ten determinations. Thus the errors on stellar parameters and  metallicities of the GIRAFFE stars are likely underestimated.    

\begin{figure}
   \centering
  \includegraphics[width=0.5\textwidth]{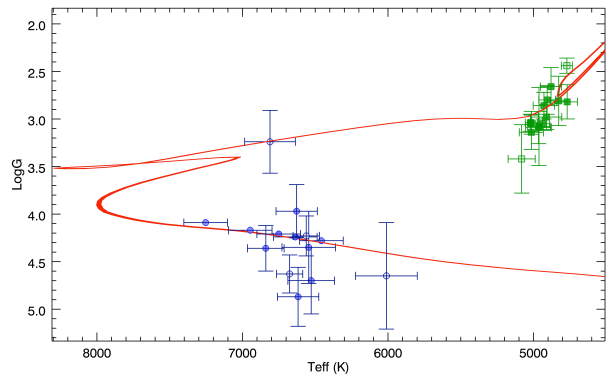}
    \caption{The HR diagram with the PARSEC isochrone for an age of 0.98 Gyr and a metallicity Z=0.025 (red curve) and the recommended Gaia-ESO parameters of  UVES stars (green squares) and 
    of   GIRAFFE stars (blue circles). The stars that are members on the basis of their radial velocity but with [Fe/H] outside 1-$\sigma$ from the cluster average metallicity are shown with empty symbols.} 
         \label{Fig:hr}
   \end{figure}

In Fig.~\ref{Fig:hr} we present the HR diagram of the candidate member stars based on RV with a PARSEC  isochrone  \citep{parsec} 
suitable for the age and metallicity 
of Be~81 (see next Section). We show with different colours the  UVES and  GIRAFFE stars. 
The most metal poor  UVES star is located in the upper part of the plot, quite far away from the location of the other
clump stars. The most metal rich star is located just below the main locus of the clump stars. 
The  GIRAFFE targets were selected among main sequence (MS) stars.  Several members follow, within the errors, the PARSEC isochrone, 
while other stars have very uncertain parameters and are not in agreement with the isochrone.

\section{The fundamental parameters of Be~81}

\subsection{The chemical properties of Be~81}
\label{sec_chem}

\begin{figure*}
   \centering
  \includegraphics[width=1.00\textwidth]{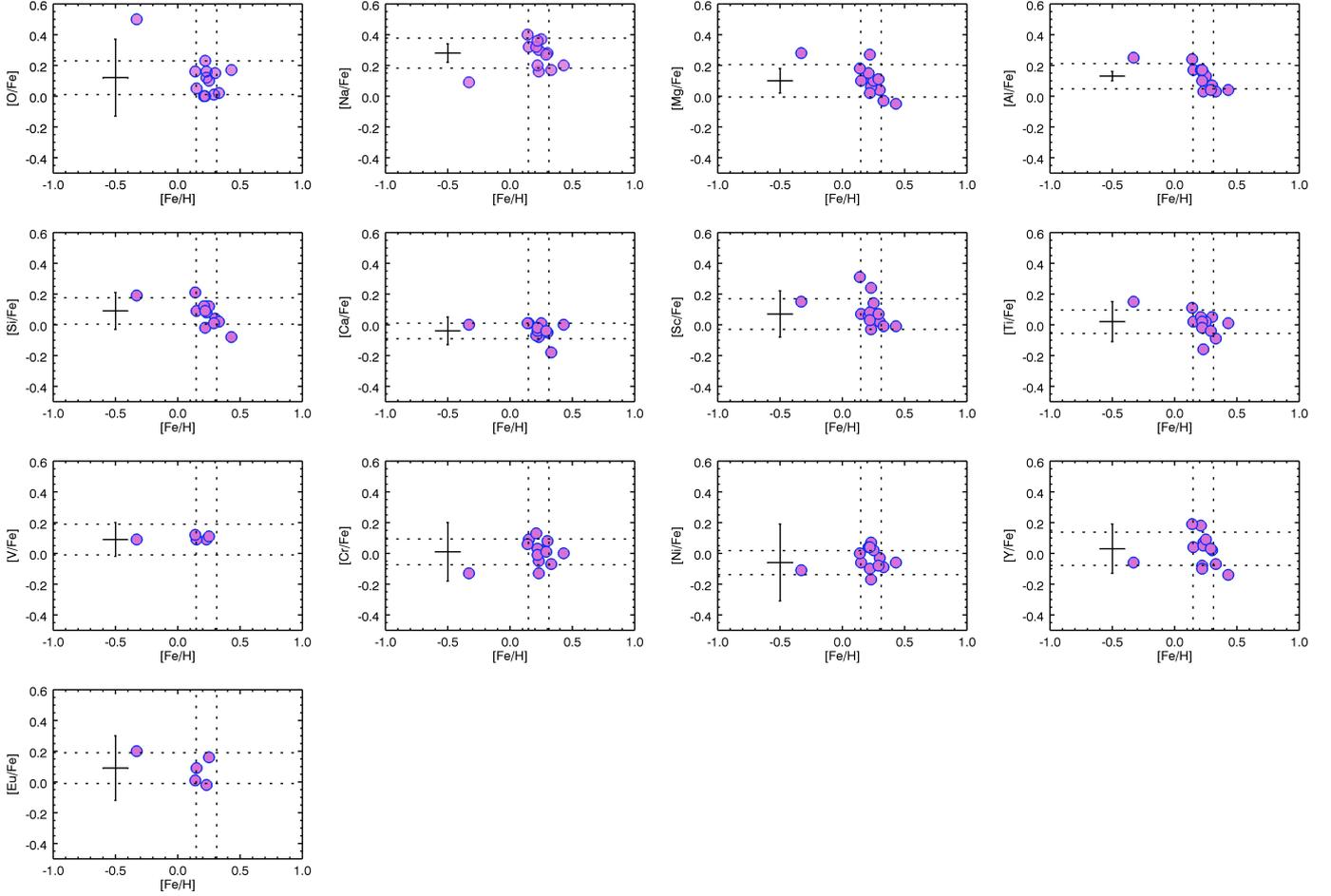}
    \caption{Abundance ratios of stars in Be~81. The dotted lines indicate the median$+\sigma$ and median$-\sigma$ regions in both [El/Fe] and [Fe/H]. 
The typical 
    error bar on each star are shown in the left side of each panel.}
   \label{Fig:abu}
         \end{figure*}

The high-resolution and high-quality  UVES spectra allow WG 11 to determine the detailed composition of the stars of Be~81. 
In Table~\ref{tab_abu} we report the elemental abundances of the  UVES stars.
These are graphically presented in Fig.~\ref{Fig:abu}, where we show [El/Fe] of each individual  UVES star. Their median values, computed excluding the metal-poor non-member star are shown in Table~\ref{tab_meanabu}. 
The abundances are scaled consistently to the Solar abundances obtained 
in the analysis of WG~11. The adopted iron abundances are those listed in the [Fe/H] column of {\sc idr2/3} table of recommended parameters\footnote{[Fe~I/H] and [Fe~II/H] abundances are available as well in the {\sc idr2/3} table, but not for all stars of each cluster.}. 
In Table~\ref{tab_solar} we show the atmospheric parameters and elemental abundances obtained 
for the selected Solar spectrum. 
For elements without a WG~11 determination, we scaled with the Solar abundance of \citet{grevesse07}. 

The abundance distribution interestingly confirms
the compactness of the cluster in most abundance ratios, whereas the most metal-poor star has a different chemical pattern. 
This is demonstrated in Fig.~\ref{Fig:abu}, which shows the dispersions around the median values (dashed lines) computed with the Robust sigma  IDL routine and indicated with $\sigma$ in Table~\ref{tab_meanabu}, as well as 
the typical error bar (black crosses in the left part of each panel and indicated with $<\delta>$) computed as the average of the individual stellar errors.
The error on the parameters and abundances of each star are discussed in \citet{smi14}, where they have been estimated with the method-to-method dispersion.
For most elements these two quantities are comparable indicating a high degree of chemical homogeneity of the cluster, as already noticed for NGC~6705, Tr~20 and NGC~4815 with {\sc idr1} data by \citet{magrini14}.
Be~81 is slightly enhanced in some of the $\alpha$-elements and in one of the neutron capture elements, having [O/Fe], [Mg/Fe], [Si/Fe] and [Eu/Fe] $>$0.1~dex, while it has [Ti/Fe] and [Ca/Fe] close to zero.  
In this context, the most metal poor star is clearly distinguishable not only for its lower [Fe/H] but also for differing more than 1-$\sigma$ in several elemental abundances, such as [O/Fe], 
[Na/Fe], [Mg/Fe], [Al/Fe], [Si/Fe], [Ti/Fe], [Cr/Fe], and [Eu/Fe]. Taken altogether these chemical signatures indicate a clear non-membership of 19015978-0028183, 
while the most metal rich star (19013651-0027021) is consistent within 1-$\sigma$ in all panels with the median values of the cluster and thus can be considered a {\em bona fide} member star.

\begin{table*}[h!]
\begin{center}
\caption{Abundances in the 12+$\log$(X/H) form of the  UVES stars in Be~81 from {\sc idr2/3} recommended values. }
\begin{tabular}{clllllllllllllllllllll}
\hline\hline
{\sc Cname}     &          O~{\sc i}  &  Na~{\sc i}                   &Mg~{\sc i}                  & Al~{\sc i}                 &Si~{\sc i}                      &Ca~{\sc i}                & Sc~{\sc i}                                  \\
\hline\hline
  19013537-0028186  &  9.11$\pm$0.05 & 6.86$\pm$0.31 & 8.05$\pm$0.29 & 6.82$\pm$0.09 & 7.80$\pm$0.12  &6.54$\pm$0.13 & 3.54$\pm$0.04        \\
  19013631-0027447  &  9.01$\pm$0.05 & 6.81$\pm$0.07 & 8.05$\pm$0.08 & 6.83$\pm$0.02 & 7.81$\pm$0.15  &6.45$\pm$0.08 & 3.69$\pm$0.15   \\    
  19013651-0027021  &  9.26$\pm$0.05 & 6.91$\pm$0.05 & 8.09$\pm$0.15 & 6.92$\pm$0.03 & 7.81$\pm$0.18  &6.72$\pm$0.10 & 3.64$\pm$0.17                       \\
  19013910-0027114  &  8.86$\pm$0.05  &6.75$\pm$0.06 & 7.96$\pm$0.02 & 6.77$\pm$0.06 & 7.70$\pm$0.09  &6.45$\pm$0.10 & 3.44$\pm$0.17   \\    
  19013997-0028213  &  9.05$\pm$0.05  &6.67$\pm$0.26 & 8.00$\pm$0.17 & 6.71$\pm$0.19 & 7.77$\pm$0.14  &6.44$\pm$0.07 & 3.42$\pm$0.15  	\\ 
  19014004-0028129  &  9.01$\pm$0.05  &6.90$\pm$0.06 & 8.06$\pm$0.01 & 6.83$\pm$0.06 & 7.83$\pm$0.11  &6.55$\pm$0.10 & 3.61$\pm$0.08   \\    
  19014127-0026444  &  9.01$\pm$0.05  &6.78$\pm$0.35 & 8.01$\pm$0.10 & 6.81$\pm$0.03 & 7.81$\pm$0.12  &6.44$\pm$0.13 & 3.54$\pm$0.20                  \\
  19014194-0028172  &  8.87$\pm$0.05  &6.81$\pm$0.05 & 8.07$\pm$0.13 & 6.83$\pm$0.01 & 7.79$\pm$0.11  &6.43$\pm$0.07 & 3.49$\pm$0.12             \\
  19014228-0027388  &  8.88$\pm$0.05  &6.86$\pm$0.06 & 8.20$\pm$0.01 & 6.84$\pm$0.01 & 7.77$\pm$0.12  &6.47$\pm$0.07 & 3.52$\pm$0.19                  \\
  19014498-0027496  &  9.11$\pm$0.05  &6.70$\pm$0.29 & 7.95$\pm$0.09 & 6.77$\pm$0.02 & 7.66$\pm$0.13  &6.49$\pm$0.09 & 3.47$\pm$0.21                      \\
  19014525-0023580  &  8.96$\pm$0.05  &6.81$\pm$0.13 & 7.96$\pm$0.14 & 6.83$\pm$0.09 & 7.75$\pm$0.09  &6.50$\pm$0.07 & 3.46$\pm$0.08  \\    
  19014769-0025108  &  8.96$\pm$0.05  &6.82$\pm$0.04 & 8.03$\pm$0.07 & 6.83$\pm$0.03 & 7.81$\pm$0.19  &6.44$\pm$0.05 & 3.67$\pm$0.11  \\    
  19015261-0025318  &  8.96$\pm$0.05  &6.84$\pm$0.29 & 8.11$\pm$0.01 & 6.78$\pm$0.08 & 7.76$\pm$0.12  &6.54$\pm$0.09 & 3.58$\pm$0.15                  \\   
  19015978-0028183  &  8.83$\pm$0.05  &6.04$\pm$0.06 & 7.66$\pm$0.01 & 6.37$\pm$0.07 & 7.32$\pm$0.08  &5.96$\pm$0.10 & 3.04$\pm$0.11   \\ 
 \hline\hline 
  {\sc Cname}               &Ti~{\sc i}                              & V~{\sc i} & Cr~{\sc i}                & Ni~{\sc i}         &Y~{\sc ii}  		 &Eu~{\sc ii}  \\
\hline\hline
  19013537-0028186  &5.27$\pm$0.18    &-                   &  5.99$\pm$0.14  &6.52$\pm$0.29  &2.53$\pm$0.12 &  	- \\
  19013631-0027447  &5.18$\pm$0.11   &4.21$\pm$0.11 &  5.79$\pm$0.22  &6.55$\pm$0.25  &2.51$\pm$0.16 &0.73$\pm$0.26\\    
  19013651-0027021  &5.36$\pm$0.17    &-   &  6.04$\pm$0.17  &6.62$\pm$0.32  &2.50$\pm$0.22 &- \\
  19013910-0027114  &5.09$\pm$0.11    &4.13$\pm$0.11&  5.85$\pm$0.23  &6.34$\pm$0.21  &2.40$\pm$0.32 &0.76$\pm$0.13\\    
  19013997-0028213  &4.99$\pm$0.16   &-   		&  5.71$\pm$0.23  &6.31$\pm$0.34  &2.49$\pm$0.12 &-  \\ 
  19014004-0028129  &5.19$\pm$0.09   &4.25$\pm$ 0.11&  5.88$\pm$0.19  &6.52$\pm$0.26  &2.55$\pm$0.16 &0.93 $\pm$0.21\\    
  19014127-0026444  &5.16$\pm$0.13    &-        &  5.87$\pm$0.16  &6.49$\pm$0.19  &2.47$\pm$0.23 &-\\
  19014194-0028172  &5.18$\pm$0.13   &-              &  5.95$\pm$0.23  &6.50$\pm$0.24  &2.60$\pm$0.19 &- \\
  19014228-0027388  &5.16$\pm$0.18    &-         &  5.86$\pm$0.20  &6.51$\pm$0.25  &2.35$\pm$0.14 &- \\
  19014498-0027496  &5.12$\pm$0.11    &-     &  5.82$\pm$0.16  &6.37$\pm$0.29  &2.33$\pm$0.22 &-  \\
  19014525-0023580  &5.17$\pm$0.11   &4.18$\pm$ 0.12&  5.84$\pm$0.17  &6.46$\pm$0.22  &2.44$\pm$0.28 &0.69$\pm$0.06\\    
  19014769-0025108  &5.17$\pm$0.13    &4.15$\pm$ 0.17&  5.81$\pm$0.20  &6.39$\pm$0.24  &2.54$\pm$0.11 &0.67$\pm$0.10\\    
  19015261-0025318  &5.17$\pm$0.10    &-        &  5.91$\pm$0.16  &6.46$\pm$0.25  &2.53$\pm$0.19 &-\\   
  19015978-0028183  &4.74$\pm$0.10  &3.65$\pm$ 0.22&  5.15$\pm$0.13  &5.81$\pm$0.18  &2.32$\pm$0.26 &0.39$\pm$0.27 \\ 
\hline \hline
\end{tabular}
\label{tab_abu}\\
\end{center}
\end{table*}

\begin{table}
\begin{center}
\caption{Median Elemental abundances with their dispersions and average errors. The non-member star is excluded from the median.   }
\tiny
\begin{tabular}{lrll}
\hline\hline
Element & Abundance & $\sigma$ & $<\delta>$ \\
\hline\hline
$[$Fe/H$]$    & 0.23 &0.08 & 0.10  \\
$[$O/Fe$]$   & 0.12&0.11& 0.05 \\
$[$Na/Fe$]$ & 0.30&0.10& 0.06      \\
$[$Mg/Fe$]$ & 0.10&0.08&0.03      \\
$[$Al/Fe$]$   & 0.15&0.10&0.08      \\
$[$Si/Fe$]$   & 0.09&0.09&0.12      \\
$[$Ca/Fe$]$  &-0.02&0.05&0.09     \\
$[$Sc/Fe$]$ &  0.07&0.10&0.15       \\
$[$Ti/Fe$]$    & 0.02&0.08&0.13     	\\
$[$V/Fe$]$    & 0.09& 0.01&0.11     \\
$[$Cr/Fe$]$   & 0.02& 0.08&0.19    \\
$[$Ni/Fe$]$   & -0.03& 0.08&0.25    \\
$[$Y/Fe$]$  & 0.03& 0.10&0.16	      \\
$[$Eu/Fe$]$  &  0.09& 0.10& 0.21   \\
\hline \hline
\end{tabular}
\label{tab_meanabu}
\end{center}
\end{table}

\begin{table}
\begin{center}
\caption{{\sc idr2/3} Solar parameters and abundances.  }
\tiny
\begin{tabular}{lll}
\hline\hline
  T$_{\rm eff}$    &  $\log$~g  & $\xi$  \\
   (K)	&			& km~$s^{-1}$\\
\hline
 5777$\pm$35   &4.43$\pm$0.13 & 1.04$\pm$0.16 \\
\hline \hline
Element & Abundance ({\sc idr2/3}) & Abundance (G07)  \\
\hline
{\sc F}e& 7.53$\pm$0.06 & 7.45$\pm$0.05   \\
{\sc O}   &  -  & 8.66$\pm$0.05 \\
{\sc N}a& 6.28$\pm$0.02& 6.17$\pm$0.04  \\
{\sc M}g & 7.71$\pm$0.06&7.53$\pm$0.09\\
{\sc A}l   & 6.45$\pm$0.04&6.37$\pm$0.06\\
{\sc S}i   & 7.46$\pm$0.04&7.51$\pm$0.04\\
{\sc C}a &6.29$\pm$0.03&6.31$\pm$0.04\\
{\sc S}c &  3.22$\pm$0.04&3.17$\pm$0.10 \\
{\sc T}i    & 4.98$\pm$0.02&4.90$\pm$0.06\\
{\sc V}    & 3.89$\pm$0.05&4.00$\pm$0.02\\
{\sc C}r   & 5.61$\pm$0.06&5.64$\pm$0.10\\
{\sc N}i & 6.15$\pm$0.14&6.23$\pm$0.04\\
{\sc Y}  & 2.08$\pm$0.11&2.21$\pm$0.02 \\
{\sc E}u  &  -& 0.52$\pm$0.06\\
\hline \hline
\end{tabular}
\label{tab_solar}
\end{center}
\tiny{Abundances expressed in 12$+\log$(El/H) form. }
\end{table}

\subsection{Age, distance and reddening of Be~81}

Thanks to accurate metallicity measurements  and  membership based on RV determinations, we are able to derive an improved age, distance, and average reddening for Be~81 by using the classical approach of isochrone fitting. As done for the open cluster Trumpler~20 \citep[see][]{donati14GES}, we chose four different sets of isochrones to take into account how the results change by adopting different models and  quantify the effect of different input physics on the derived parameters:  PARSEC \citep{parsec},  BASTI \citep{basti}, Dartmouth \citep{dart}, and  Victoria-Regina \citep{victoria}. We adopted the $B$, $V$, and $I$ photometric catalogue described in \cite{donati14}.
The accurate spectroscopic analysis of the cluster metallicity from the UVES spectra clearly indicates a super-solar chemical mixture for Be~81, in contrast to previous findings which were based only on Ca-triplet spectra.
Using this information, we are able to select a more appropriate set of isochrones and thus to  
obtain a more accurate determination of the cluster evolutionary status, reddening, and distance modulus.

The best-fitting isochrone is chosen by visual examination as
the one which can describe at the same time the main age-sensitive evolutionary phases: the luminosity and colour of the
main sequence turn-off (MSTO), red hook (RH), and red clump (RC), defined in \citet{donati14}, in two different photometric colours 
($B-V$ and $V-I$, adopting $R_V=3.1$, $A_B/A_V=1.29719$, and $A_I/A_V=0.60329$ taken from \citet{cardelli89} and \citet{odonnell94}). We started by using the metallicity resulting from the  UVES spectroscopic estimate, i.e. [Fe/H]$=+0.23$ dex (see Section~2.2), and we converted it to Z taking into account the different solar abundances of the four sets of isochrones.  The difference in luminosity between the MSTO and RC is adopted as the most reliable constraint on the cluster age. In fact, while the colour difference between the same two indicators is related to age as well, it is used here as secondary age indicator because colour properties are more affected by theoretical uncertainties, like colour transformations and the super-adiabatic convection, while luminosity constraints are more reliable. The errors on the estimated cluster parameters are mainly due to the uncertainties in the definition of the age indicators because of the heavy contamination by field interlopers, while the photometric error does not have a significant impact on the error budget \citep[see][]{donati14}. We  make use of the RV to select the most probable cluster members among the many stars observed in order to highlight the most probable cluster sequences in the CMD and to have more stringent constraints on the best-fitting models. Moreover, we tweak the age determination looking for the best-fitting isochrone in the theoretical plane $T_{\rm eff}$, $\log$ g by using the RC star parameters of the  UVES targets (see Table~\ref{tab_par}).

\begin{figure*}
\begin{center} \includegraphics[scale=0.90]{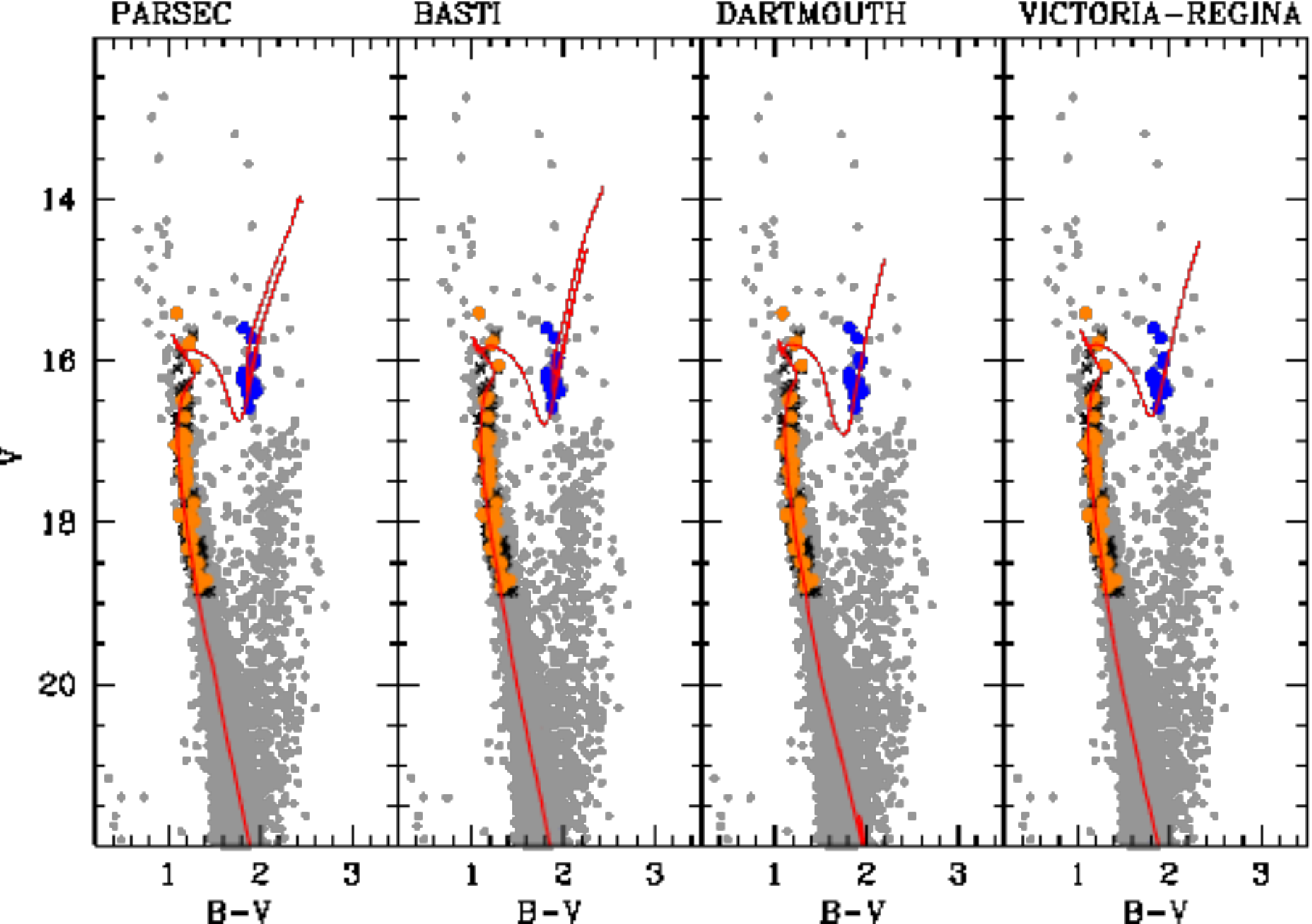} \caption{\label{fig:isoc}CMD obtained for 
stars inside 3$\arcmin$ using the photometry from \cite{donati14} and best-fit isochrone for different evolutionary models (from left to right: PARSEC, BASTI, DARTMOUTH, and Victoria-Regina). GES target non members 
are plotted as black crosses, while members are highlighted with orange (GIRAFFE) and blue (UVES) 
points. See Table \ref{tab_parisoc} for the adopted parameters for the 
isochrone fitting.} 
\end{center}
\end{figure*}

For the PARSEC set (see Fig.~\ref{fig:isoc}, left panel), for which Z$_{\odot}=0.015$, we adopted Z=0.025. The luminosity and colour of the RC are well 
reproduced as are those of the RH and upper MS. In Fig.~\ref{Fig:hr} we show, as an example, the isochrone fit in the theoretical plane which, together with the photometric data, 
was used to determine the best age, i.e. 1.0$\pm$0.1~Gyr. We note that in the case of the $V,V-I$ CMD the RC and MS colours of the model are slightly bluer than observations, and this is evident also in the best-fit with the three other isochrones. As mentioned above, this effect may be due to uncertainties in the temperature-colour transformations or  to the differences in chemical mixtures between models and observations or, finally, to small systematics in the $B$ and $I$ photometry that in principle might have a larger impact on colours, of the order of some hundredths of magnitude. Considering, however, that metallicity is fixed by spectroscopy and age by the difference in luminosity between RC and TO, we actually used $V,B-V$ and $V,V-I$ photometry to adjust the distance modulus and reddening, and deemed this fit quite satisfying and the small mismatch of the RC colour acceptable.
In the case of the BASTI isochrones (Z=0.03), the best-fit is obtained for a younger age of $0.75\pm0.1$ Gyr. 
The Dartmouth isochrones and the Victoria-Regina isochrones do not include the evolved phases after the RGB for the age 
of Be~81 and thus the cluster's parameters for these two models are more loosely constrained. We proceeded by choosing 
the best-fitting solution as the one that best matches the MS and MSTO morphologies (see Fig.~\ref{fig:isoc}, right panel).
We used models with [Fe/H]=$+$0.23 for both sets of isochrones. 
The results obtained with the four sets are given in Table~\ref{tab:isoc}, where we indicate
age, distance modulus, reddening, distance from the Sun and the Galactocentric distance from the Galactic plane, 
and mass of the stars at the MSTO. The cluster's centre coordinates are taken from \citet{donati14}: RA=19:01:42.8 and Dec=-0:27:07.7.

Since from spectroscopic analysis it turned out that Be~81 stars show an $\alpha$-enhanced chemical pattern ([$\alpha$/Fe]$\sim$+0.2, see Section~\ref{sec_chem}), we look also at the model predictions when an $\alpha$-enhanced mixture is taken into account. Slightly younger ages are obtained.  The results are reported in Table~\ref{tab:isoc}. 

\begin{figure*}
\begin{center} \includegraphics[scale=0.90]{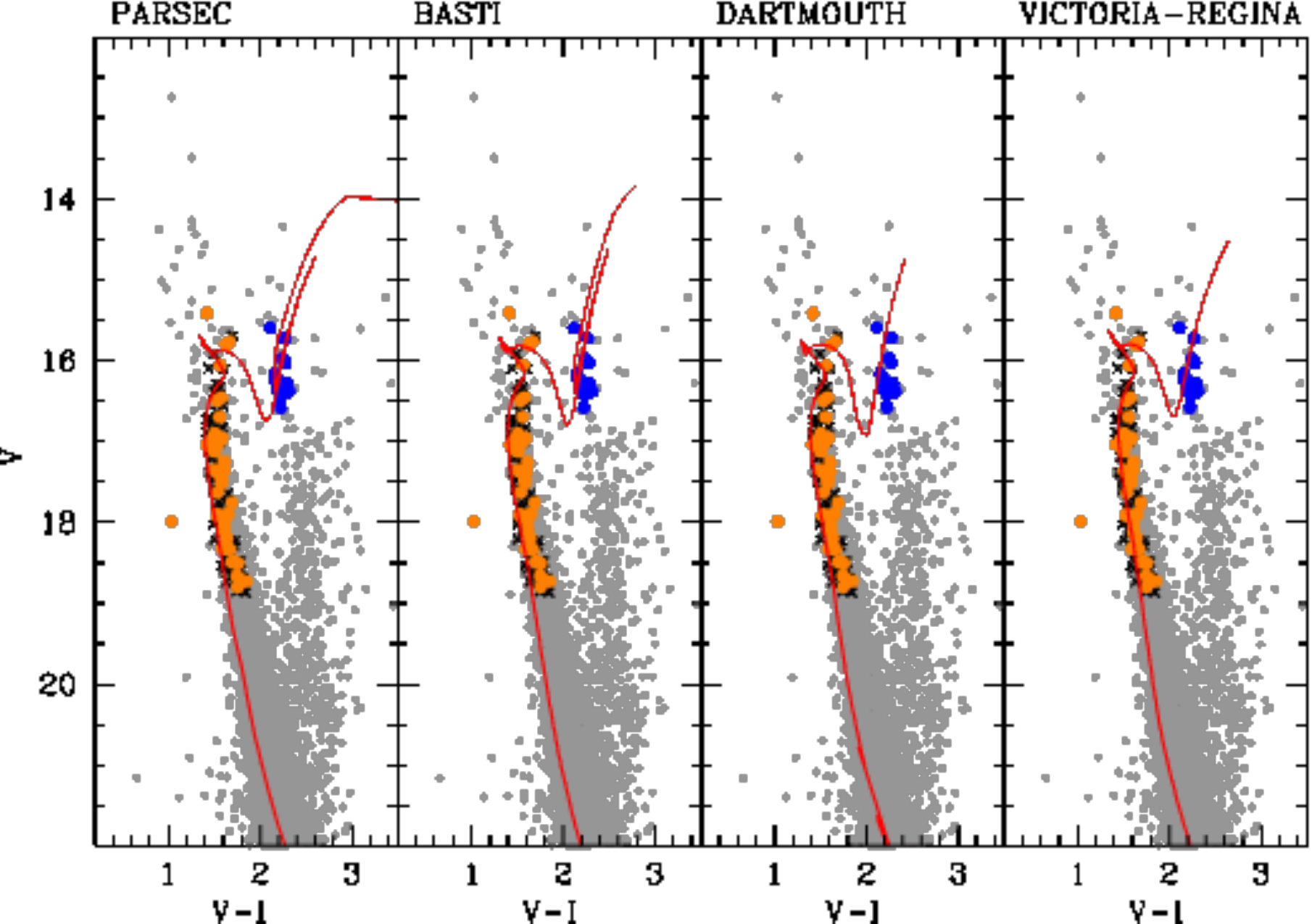} \caption{\label{fig:isoc2}The same as Figure~\ref{fig:isoc} but in the $V$,$V-I$ CMD} 
\end{center}
\end{figure*}

\begin{table*}
\centering
\caption{ \label{tab_parisoc} Results, errors, and estimated systematic uncertainties using different evolutionary models with [Fe/H]$\sim$0.2.}
\begin{tabular}{lccccccc}
\hline
\hline
 Model & age & $(m-M)_0$ & $E(B-V)$ & $d_{\odot}$ & $R_{GC}$ & z & $M_{TO}$ \\
       & (Gyr) & (mag) & (mag) & (kpc) & (kpc) & (pc) & ($M_{\odot}$)\\
\hline \hline
PARSEC        & $0.98\pm$0.1  & $12.72\pm$0.1 & $0.83\pm$0.02 & 3.50 & 5.45 & -152.18 & 2.2\\
BASTI         & $0.75\pm$0.1  & $12.85\pm$0.1 & $0.86\pm$0.02 & 3.72 & 5.33 & -161.57 & 2.3\\
BASTI$^a$     & $0.65\pm$0.1  & $12.75\pm$0.1 & $0.89\pm$0.02 & 3.55 & 5.42 & -154.30 & 2.3\\
Dartmouth     & $0.90\pm$0.15 & $12.72\pm$0.1 & $0.84\pm$0.04 & 3.50 & 5.45 & -152.18 & 2.2\\
Dartmouth$^a$ & $0.80\pm$0.15 & $12.70\pm$0.1 & $0.83\pm$0.04 & 3.47 & 5.47 & -150.79 & 2.2\\
Victoria-Regina      & $0.90\pm$0.15 & $12.75\pm$0.1 & $0.85\pm$0.04 & 3.50 & 5.42 & -154.30 & 2.3\\
\hline\hline
\end{tabular}\\
$^a \alpha$-enhanced models: [$\alpha$/Fe]$\sim+0.4$ dex in the case of BASTI (we chose the models with [Fe/H]=0.16 dex, the closest metallicity to the spectroscopic derivation) and [$\alpha$/Fe]=+0.2 dex ([Fe/H]=0.23 dex) in the case of Dartmouth.  
\label{tab:isoc}
\end{table*}

With respect to literature values \citep[see][]{donati14, sagar98}, we found an acceptable agreement with the determination of the age, distance and reddening  of the cluster.

\section{A comparison with inner disc clusters and field stars} 

\begin{table*}
\begin{center}
\caption{Clusters' parameters}
\tiny
\begin{tabular}{lrlllllll}
\hline\hline
Name 	&RA 	&dec 	& l	&b	&$E(B-V)$ 	&Age 	&R$_{\rm GC}$(a) & Ref.	\\
		& \multicolumn{2}{c}{J2000.0} &deg.	&deg.      &                       &(Gyr) 	&(kpc) 	 \\		
\hline\hline
Be~81		&19:01:36	 &	-00:31:00 &	34.505 &	-2.07 &      0.85 & 0.98$\pm$0.1  & 5.45    & present work \\
NGC 6705 	&18:51:05 &	-06:16:12 &	27.31  &	-2.78 &	0.43 	&0.30$\pm$0.05 & 6.3      &{\sc idr1},  \citet{cantat14}\\
Tr~20		&12:39:32 &	-60:37:36 &	301.48& 	  +2.22 &	0.33 	&1.50$\pm$0.15 & 6.88 	         &{\sc idr1},  \citet{donati14GES}\\
NGC 4815 	&12:57:59 &	-64:57:36 &	303.63& 	-2.10 &	0.72 	&0.57$\pm$0.07 & 6.9 	 &{\sc idr1},  \citet{friel14}\\
\hline \hline
\end{tabular}
\label{tab_clu_par}
\end{center}
\tiny{
(a) computed with R$_{\odot}$= 8 kpc.}
\end{table*}

The large variety of available chemical elements, belonging to different nucleosynthesis channels, allows us to compare,  in a fully consistent way, the detailed chemical patterns of open clusters and of field stars observed and analysed homogeneously by the Gaia-ESO Survey. 
Besides Be~81, three other old/intermediate-age clusters are available in {\sc idr2/3}. 
For all of them, namely Tr~20, NGC~6705, and NGC~4815, the stellar parameters, abundances, and radial velocities,  
were also provided in the first data release, {\sc idr1}, and analysed in several papers: 
\citet{donati14GES} derived membership, age, distance, and reddening of Tr~20; \citet{friel14} derived the parameters of NGC~4815 and studied its chemical composition; 
\citet{cantat14} studied the properties of NGC~6705 and searched for possible inhomogeneities in its chemical composition; finally \citet{magrini14}
searched for unique signatures in the chemical patterns of the three clusters as possible indication of their birthplace.
Their CNO abundances have been discussed by \citet{tautvasiene15} using {\sc idr2/3}  results. 
Their main parameters are summarised in Table~\ref{tab_clu_par}.
\begin{figure*}
   \centering
  \includegraphics[width=0.95\textwidth]{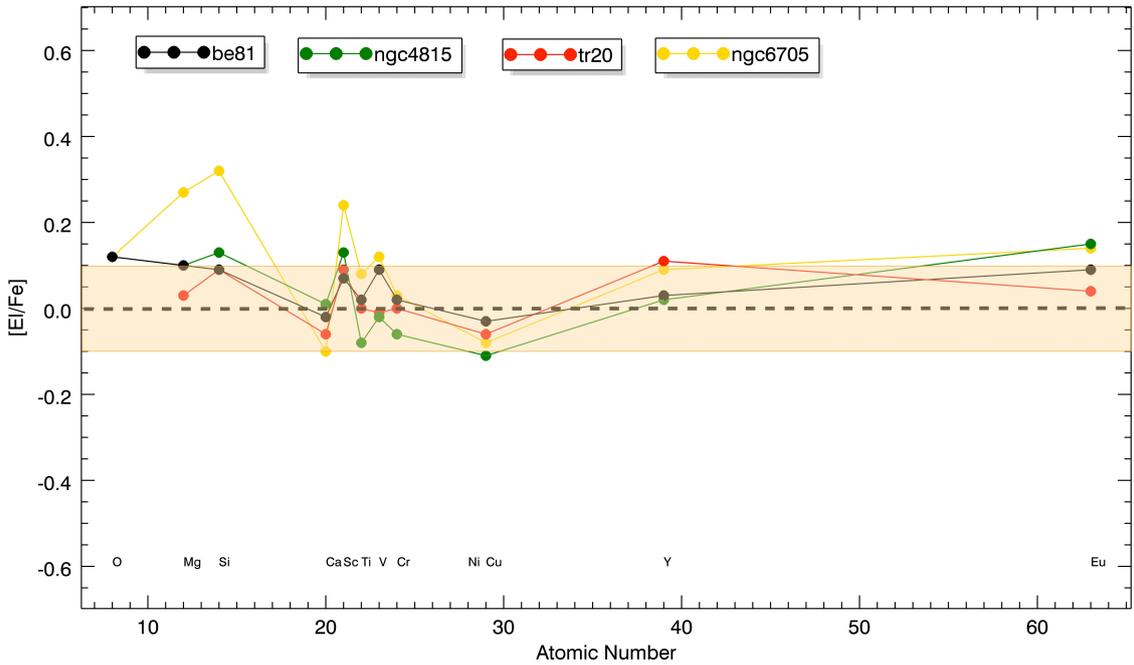}
    \caption{The abundance ratios of the {\sc idr2/3} old and intermediate-age open clusters located in the inner disc as a function of the atomic number. 
    In black are the abundance ratios of Be~81, in yellow  NGC~6705, in green NGC~4815, and in red Tr~20.  The dashed line marks the solar abundance, and the shaded area the 1-$\sigma$ regions, with $\sigma$=0.1~dex. 
              }
         \label{Fig:cpbe81}
   \end{figure*}

\begin{figure*}
   \centering
  \includegraphics[width=0.95\textwidth]{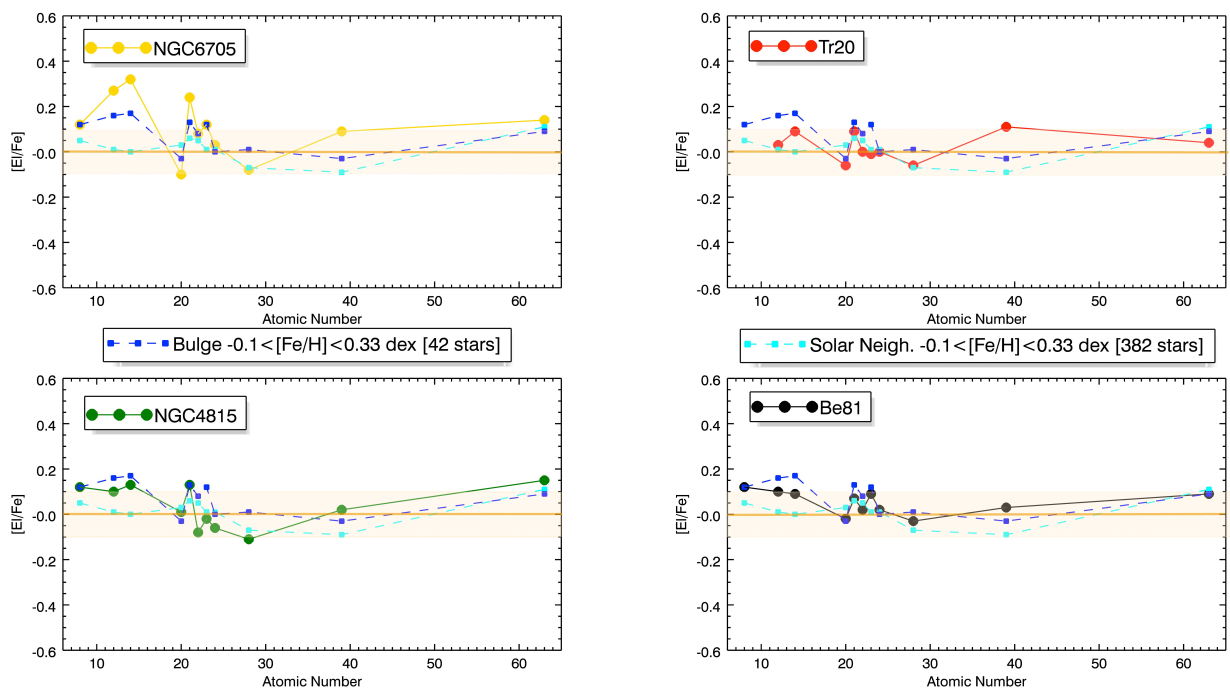}
    \caption{The abundance ratios vs atomic numbers of the {\sc idr2/3} stars in the solar neighbourhood and in the inner disc/bulge (selected in the [Fe/H] range from $-$0.10 to $+$0.33~dex, i.e., including the metallicity ranges 
    of the four clusters) compared with the chemical pattern of NGC6705, Trumpler~20, NGC~4815, and Be~81. The continuous lines mark the Solar abundances, and the shaded area the 1-$\sigma$ 
    region, with $\sigma$=0.1~dex. 
              }
         \label{Fig:cpdisk}
   \end{figure*}

Here we use the  {\sc idr2/3} stellar parameters and abundances of NGC~4815, NGC~6705, and Tr~20 to compare their chemical patterns with  Be~81 and with  field stars. 
Although shown in Fig.~\ref{Fig:abu}, in the following we do not discuss Al and Na in terms of Galactic chemical evolution. They are indeed potentially interesting elements for comparison with the field samples and deduction of the chemical evolution of the inner disc. However, the abundances derived in {\sc idr2/3} are all computed in local thermodynamic equilibrium (LTE), and those elements are affected by relevant non-local thermodynamic equilibrium (NLTE) effects. 
All stars observed with  UVES in Be~81 are in the same evolutionary phase (RC), so that the  internal dispersion in abundance is not affected  by this problem. However, the absolute value of   Na and Al abundances and the comparison with stars of different parameters could be unreliable without these corrections. Furthermore, both Na and Al are subject to (extra-)mixing processes in stars of the mass involved here \citep[see, e.g.,][]{cl10,kl14}. These elements will be  discussed in a forthcoming paper (Smiljanic et al., in prep) and will not be used. 

As is well known, the ratio of $\alpha$-elements to iron, [$\alpha$/Fe], is commonly used to trace the star-formation timescale in a system, 
because it is sensitive to the ratio of SNII (massive stars) to SNIa (intermediate mass binary systems with mass transfer) that have occurred in the lifetime of the system. 
However, the so-called $\alpha$-elements, that are generally associated with the explosion of massive stars,  do not have  exactly the same place of production, and some of them, as Si, Ca, and Ti, have also a significant 
contribution from SNIa. 
In addition, there are also differences in the way these elements are produced in massive stars: %
O and Mg are produced during the hydrostatic He burning, and their yields are not expected to be significantly affected by the SNII explosion conditions, while Si, Ca and Ti are mostly produced during the SNII explosion. This distinction is also seen in the observations \citep[e.g.,][]{ful07}, where Si, Ca and Ti usually track one another, but O and Mg often show different trends with [Fe/H]. 
The "pure" $\alpha$-elements (O, Mg) are produced almost entirely by massive stars, and thus in shorter time-scales. 
In the bulge, the observed Mg overabundances and the general $\alpha$-enhancement agree with the predictions of  the chemical evolution models of 
\citet{matteucci90} and \citet{grieco12} and they are explained with a rapid formation time scale for the bulge. 
A summary of the place of production and of the contribution of SNII and SNIa  is shown in Table~\ref{tab:el}. 
The Table is computed using the information given in \citet{woosley02} and in  \citet[][model W7]{iwamoto99} and summarises the origin of the main isotope of each element. 

In Figure~\ref{Fig:cpbe81} we show the median 
[El/Fe] abundance ratios versus the atomic number
in the four old/intermediate-age clusters located within 7~kpc from the Galactic Centre, whereas
in Figure~\ref{Fig:cpdisk} 
we show the chemical pattern of the four clusters 
compared to the median abundance ratios of stars with $-$0.10$<$[Fe/H]$<+$0.33~dex in the solar neighbourhood  [382 stars] and 
in the inner-disc/bulge [42 stars] samples, i.e., embracing the maximum metallicity range of the  four clusters within 1-$\sigma$.
We have excluded all stars with errors on [Fe/H] larger than 0.15~dex.  

In both figures, the available elements include the $\alpha$-elements (O, Mg, Si, Ca, Ti), 
the refractory elements (Sc, V),  the iron-peak elements (Cr, Ni), and the neutron capture elements (Y,  Eu).   
From Fig.~\ref{Fig:cpbe81}, we can make some general considerations on the similarities in the chemical patterns of the inner clusters. 

Considering the errors that affect the abundance ratios, we define  
an element as enhanced with respect to Solar if [El/Fe]$\geq$0.1~dex. Also,
we consider that the abundance of an element is globally enhanced 
in the inner cluster sample if at least three of the four clusters have  [El/Fe]$\geq$0.1~dex for that element.

NGC~6705, NGC~4815, and Be~81 have enhanced abundances of the $\alpha$-elements  O, Mg, and Si, with NGC~6705 having 
the highest abundance ratios of [Mg/Fe] and [Si/Fe] while [O/Fe] does not exceed a 0.1~dex enhancement. 
The group of elements from Ca to Ni shows Solar abundance and they have similar values in the four clusters. Y and Eu are very similar in all clusters, close to Solar value for [Y/Fe] and 
slightly super-solar for [Eu/Fe]. 
From Fig.~\ref{Fig:cpdisk}, we find that the chemical pattern of Be~81 is nearly identical to that of inner-disc/bulge stars: both populations show similar $\alpha$-enhancements for the 
lower atomic number  $\alpha$ elements (O, Mg, and Si), while Ca,  Ti and  iron-peak elements, are almost solar. 
For comparison, the Solar neighbourhood stars in the same metallicity range are plotted: they all have [Mg/Fe], [Si/Fe] and   [O/Fe] close to zero.

\begin{table*}
\begin{center}
\caption{Stellar Nucleosynthesis }
\scriptsize 
\begin{tabular}{llll}
\hline\hline
{\sc Element} & {\sc Main Production Site}  & {\sc mechanism} & {\sc yield(SN~Ia/SN~II)}\\
\hline\hline
$^{16}$O & Massive Stars                & Helium burning & 8\%\\ 
$^{24}$Mg & Massive Stars		& C, Ne burnings &10\% \\
$^{28}$Si & Massive Stars		& explosive and non-explosive O burning &60\%\\
$^{40}$Ca & Massive Stars		& explosive and non-explosive O burning &67\% \\
$^{45}$Sc & Massive Stars		& C, Ne burnings, $\alpha$ and $\nu$-wind (neutrino-powered wind) &49\%\\
$^{48}$Ti  & Massive Stars and SNIa		& explosive Si burning and SNIa with He detonation&63\%\\
$^{51}$V  & Massive Stars and SNIa		& explosive Si and O burnings, SNIa with He detonation, and $\alpha$ and $\nu$&88\%\\
$^{52}$Cr  & Massive Stars and SNIa		& explosive Si burning, SNIa with He detonation, and $\alpha$ &84\%\\
$^{56}$Fe  & Massive Stars and SNIa		& explosive Si burning and SNIa &88\%\\
$^{58}$Ni  & Massive Stars (and SNIa)		& $\alpha$ ($\alpha$-rich freeze-out from nuclear statistical equilibrium) and SNIa &75\%\\
$^{50}$Y    & Massive Stars  		         & He-burning s-process,  and $\nu$-wind&--\\
$^{153}$Eu  & Massive Stars, compact binary merger 		& $\nu$-wind&--\\
\hline \hline
\end{tabular}
\label{tab:el}\\
\end{center}
\end{table*}

\section{The inner clusters and the disc evolution}

For the first time, thanks to the Gaia-ESO Survey, we have at our disposal  
a significant number of open clusters together with field stars, homogeneously analysed and for which elements from different nucleosynthesis channels
are available. 
They allow us to analyse the typical chemical patterns in different regions of the disc of our Galaxy. 
Clusters are unique tools because their distances and ages can be derived with great accuracy, allowing reliable placement
in space and time in our Galaxy. 
The family of clusters here described belongs to an important but poorly investigated region of our Galaxy: the inner disc. 

There are two main results: 
\begin{itemize}
\item[1.] As noticed in \citet{magrini14}, there are 'real' differences in the abundance patterns of the clusters even if they are located at similar distances from the Galactic centre. One very noticeable difference is the higher $\alpha$-enhancement of NGC 6705, that might suggest an inner birthplace for this cluster, or, as we see in next sections, possibly an episode of local enrichment. 
\item[2.] Despite the small but 'real' differences, a common behaviour of the family of old/intermediate-age clusters analysed so far can be seen: 
they are almost solar in most of the iron-peak and {\em heavy} $\alpha$-elements, while they are enhanced in the {\em light} $\alpha$-elements O, Mg, and Si. 
We discuss in the next sections how much this enhancement is real or due to Non-Local Thermodynamical Equilibrium (NLTE) effects.  
\end{itemize}

\subsection{Is the $\alpha$-enhancement seen  in the inner clusters real?  }

Three of the clusters of  our inner disc sample have [O/Fe], [Si/Fe], and [Mg/Fe]$\geq$0.1~dex.  [Si/Fe] is very high in NGC~6705, while it is around 0.1~dex for the 
remaining two clusters. 
First, it is necessary to investigate if the O, Mg, and Si enhancements are real or related to other effects, such as   NLTE. 
All stars analysed in the four clusters are indeed giant stars, and thus NLTE might affect the abundance of some elements. 
If the NLTE effect is large, the comparison between  
the giant stars in clusters and  the Sun or the Milky Way field solar-neighbourhood dwarf stars can be misleading.
We clarify that our 'definition' of NLTE is totally empirical: in the following we indicate as a 'NLTE effect'  the sum of all un-modelled physics plus potential model biases.

{\it Oxygen.} 
In Be~81, as well as in the other clusters and field stars, oxygen abundance was determined from the forbidden [O~I] line at 6300.31 \AA, as described in detail by \citet{tautvasiene15}. 
Lines of [O~I] are considered as very good indicators of oxygen abundances because they are not only insensitive to NLTE effects, but also give similar oxygen abundance results with 3D and 1D model atmospheres (cf. \citep{asplund04}; \citep{pereira09}). 
Thus, the comparison of giants and dwarfs should not be affected by any such bias.

{\it Magnesium.}
Mg is less  affected than the other $\alpha$-elements by model
dependent differences in the yields of SN II  
with progenitors of different masses. 
Therefore it  constitutes one of the best  
tracers of the evolutionary timescales of the Galaxy.
However, the NLTE effects in Mg are not easily quantifiable. 
From the theoretical point of view, 
\citet{merle11} computed the 
Mg~I abundance corrections for 
several lines in the optical interval as functions of atmospheric 
parameters for cool evolved stars. 
Their NLTE corrections are computed as a function of stellar parameters and metallicity. 
At solar metallicity for a star with T$_{\rm eff}$=5000~K and $\log$~g=2 the NLTE corrections of different Mg~I lines vary from -0.03 to 0.18~dex. 
From the observational point of view, several  works in the literature have analysed dwarf and giant stars in the same cluster to investigate if there is any difference in the abundance ratios related 
to the type of the  star analysed. 
A study of dwarf and giant stars in the  Hyades open cluster \citep{schuler09} showed that  Mg abundances are  overabundant by 0.2-0.5 dex in the giants relative to the dwarfs. 
Such large enhancements are not predicted by  stellar models, and they explained them as likely  due to NLTE effects, significantly larger 
than those available in the  literature. 
A similar result was obtained by \citet{yong05} comparing abundances in dwarf and giant stars in  M~67.
However, \citet{pasquini04} and by \citet{blanco15} found contrary results for the open cluster IC 4651 ([Fe/H] = 0.11).

{\it Silicon}. The few studies on its NLTE behaviour
indicate strong effects only in low metallicity stars. Recently, \citet{berg13} 
found that the effect of NLTE is a reduction of the
number densities of atoms in a minority ionisation stage, 
thus, for a given abundance,
the line equivalent width is smaller than that given by LTE. 
This effect is larger in
metal-poor stars. At solar metallicity, the opposite behaviour is encountered,  showing negative
NLTE effects such that the LTE abundance is higher than the NLTE one. 

In order to better investigate the issue
for Si and Mg, we test the nature of the $\alpha$-enhancement in our results 
using the results for one of the Gaia-ESO calibrator open
clusters (see Pancino et al. in prep.), M~67. The data for M~67 are, 
indeed, perfectly suited because they contain stars belonging to 
different evolutionary 
phases that have been analysed with the same tools and methods as the 
inner-disc clusters and field stars of {\sc idr2/3}. 

\begin{figure*}
   \centering
  \includegraphics[width=0.95\textwidth]{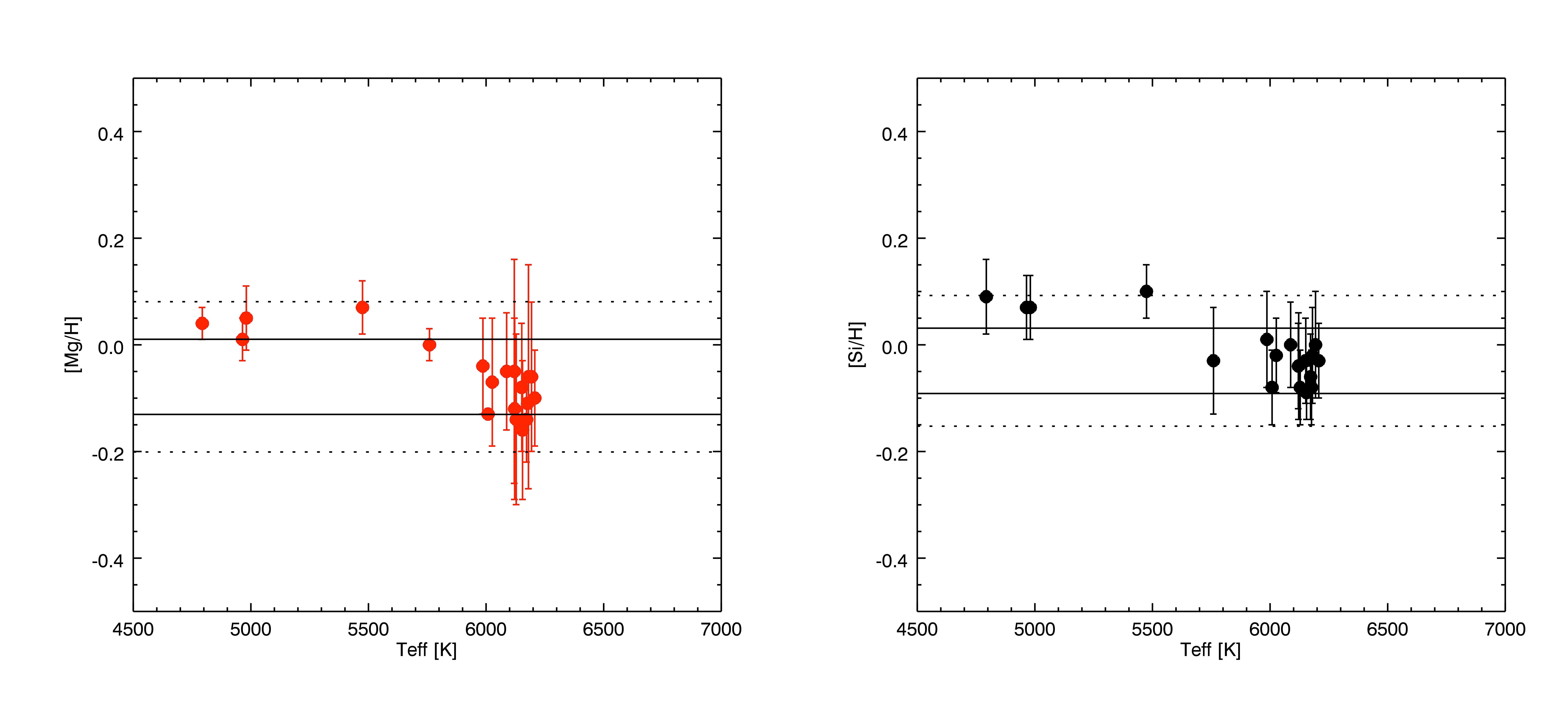}
    \caption{The abundance ratios [Si/H] and [Mg/H] vs T$_{\rm eff}$ of member stars in M~67 from  {\sc idr2/3} analysis.  The continuous lines mark the $\pm$1-$\sigma$ from the average value, whereas the 
    dotted lines indicate the $\pm$2-$\sigma$ region. 
              }
         \label{fig:m67}
   \end{figure*}

In Fig.~\ref{fig:m67}, we show [Mg/Fe] and [Si/Fe] vs T$_{\rm eff}$ in member stars of M~67 analysed in {\sc idr2/3}. 
The $\pm$1-$\sigma$ and  $\pm$2-$\sigma$ regions from the average are shown. 
The sample of stars with T$_{\rm eff} >$5700~K includes TO, MS, and SG stars. 
The stars with T$_{\rm eff} <$5700~K are instead
giant stars with  $\log$~g ranging from 2.97 to 3.46. 
We note that for both elements a slight overabundance in giants is
present with respect to TO, MS, and SG stars.
The average [Mg/H] for stars with T$_{\rm eff} <$5700~K is 0.04$\pm$0.02 and for [Si/H] is 0.08$\pm$0.01, 
while the stars with T$_{\rm eff} >$5700~K have average values of -0.09$\pm$0.04 and  -0.03$\pm$0.03 from [Mg/H] and [Si/H], respectively. 
We can use these differences to roughly estimate the amount of NLTE effects: 0.13~dex and 0.11~dex for [Mg/H] and [Si/H], respectively. If we consider the abundances with respect to iron, the amount of NLTE effects is 0.12~dex and 0.09~dex for [Mg/Fe] and [Si/Fe], respectively.


It is worth mentioning that M~67 has been used by \citet{onehag14} to 
study the effect of selective atomic diffusion of chemical elements in stars in different evolutionary phases.  
More specifically, they investigated the detailed chemical composition 
of 14 stars located on the main sequence, at the turn-off point, 
and on the early sub-giant branch, 
finding that  the heavy-element abundances are reduced in the TO and main sequence, by typically $\leq$0.05~dex, when compared to the abundances of the sub-giants (even lower for ratios of metals). We note that the effect of diffusion in our data is circumscribed
in the $\pm 1 \sigma$ area.

We finally mention that \citet{blanco15} study the differences in the abundance ratios in cluster giant and dwarf member stars. 
They show the chemical differences between dwarfs and giants in several clusters, including IC~4651, M~67, NGC~2447,  and NGC~3680. 
They find that, on average,  [Mg/Fe] and [Si/Fe] are enhanced in giant stars by $\sim$0.06~dex and $\sim$0.10~dex with respect to dwarfs. 
In particular, for M~67, the effect they measured is 0.07 and 0.10~dex in  [Mg/Fe] and [Si/Fe], respectively, thus comparable with what we find with Gaia-ESO recommended 
values.  

In conclusion, for oxygen, no NLTE has to be applied, and the three clusters for which this element has been measured show a [O/Fe]$\geq$0.1~dex.
The enhancement in [Mg/Fe] and [Si/Fe] observed in the inner-disc/bulge population and in Be~81, Tr~20, and NGC~4815 with respect to the Solar neighbourhood dwarf population  can be related to different evolutionary phases of the observed stars (see Figs.\ref{Fig:cpbe81} and \ref{Fig:cpdisk}).   
On the other hand, 
 [Mg/Fe] and [Si/Fe] are remarkably higher in NGC~6705 than in the other inner-disc clusters and in the field populations.  
We can evidently exclude that this enhancement is due solely to NLTE effects. The abundance ratios of NGC~6705 clearly indicate an $\alpha$-enhanced population whose origin needs to be investigated.

\subsection{Comparing the inner clusters with  high-$\alpha$ metal-rich population}

Once shown that the enhancements of [O/Fe], [Si/Fe] and [Mg/Fe]  in NGC~6705 are genuine, we can try to compare its chemical pattern with those of a  population that presents similar chemical characteristics, the high-$\alpha$ metal-rich stars (HAMR) with [Fe/H]=0.03$\pm$0.12.
 
The HAMR stars--whose $\alpha$-enhancement, computed with Mg, Si, and Ti, is in the range 0-0.2~dex-- are rare objects in the solar neighbourhood, and their exact nature is still under debate.
\citet{adibekyan11} identified these stars and suggested that they may form a distinct family, based on a
``gap'' in metallicity between HAMR and classical thick disc stars. Other authors also confirmed the existence of the $\alpha$-enhanced 
metal-rich stars but with different conclusions on their origin \citep[e.g.][]{gazzano-13, nissen-14, Bensby-14, Adibekyan-14}. 
A comparison of the main physical properties of HAMR stars with those from the simulations of \citet{Roskar-12}
suggests that the HAMR stars might have their origin in the central part of the Galaxy, within about 2~kpc \citep{Adibekyan-13}.
Looking at Figs.~\ref{Fig:cpbe81} and \ref{Fig:cpdisk}, one would claim that stars in Be~81 and in the inner-disc clusters are similar to the $\alpha$-enhanced metal-rich stars
discovered in the solar neighbourhood. 
They have indeed solar or super solar [Fe/H] \citep[see][and Table~\ref{tab_meanabu} for Be~81]{magrini14} and they are, in general, enhanced in the {\em light} $\alpha$-elements, as O, Mg, and Si.
At a first glance, the chemical patterns of Be~81 shown 
in Fig.~\ref{Fig:cpbe81}  pretty much resemble  the metal rich $\alpha$-enhanced population identified in the solar neighbourhood. 
Note that HAMR stars studied by \citet{adibekyan11} have T$_{eff} >$ 4900~K and only 5 stars have $\log$~g$<$3.8~dex, therefore their sample is 
essentially composed of dwarf stars. 
If confirmed, the existence of HAMR stars in an open cluster at 5-7~kpc from the Galactic Centre, might put important constraints on the stellar migration 
and in the birthplace of HAMR stars.  

However,  we should consider that  there are important differences between HAMR stars and the inner disc clusters. 
First, there is a large difference in age between the two populations.  
The age of HAMR stars in the solar neighbourhood is not precisely determined. However, they are on average believed to be quite old. 
\citet{adibekyan11}  showed that the HAMR and thick disc family stars have almost the same age, being on average older than thin disc stars by about 3 Gyr. 
At the same time, they found that HAMR stars are quite different from the thick disc stars being more metal-rich and with different chemical patterns. 
They also show properties in common with metal-rich bulge stars   \citep[e.g.,][]{ful07, Bensby-14, ness13}.
On the other hand, the age and distance of NGC~6705 are well known by isochrone fitting (see Table~\ref{tab_clu_par}),  and it is much younger. 

\begin{figure*}
   \centering
  \includegraphics[width=0.95\textwidth]{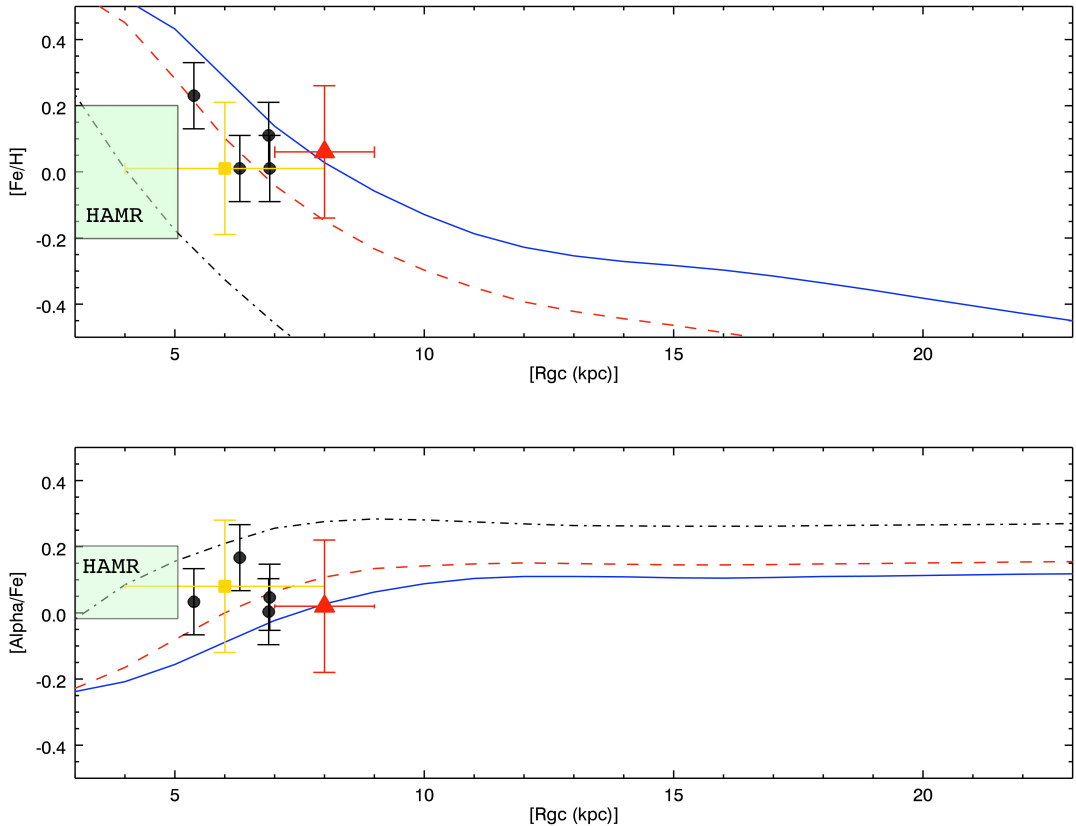}
    \caption{[Fe/H]  vs. R$_{\rm GC}$ (upper panel) and  R$_{\rm GC}$ vs. [$\alpha$/Fe] (lower panel) in the chemical evolution model of \citet{magrini09}. The three curves 
    are the curve at present time (blue), $\sim$5~Gyr ago (red) and $\sim$10 Gyr ago (black). The green shaded area indicates where and when HAMR stars might be born, the black filled circles are the four open clusters and the yellow squares are the median values of the inner disc field stars in the metallicity range from $-$0.1 to $+$0.33~dex, 
    whose [$\alpha$/Fe] are computed removing the NLTE effects. Finally, the red triangles are the median values  of the Solar neighbourhood stars in  the metallicity range from $-$0.1 to $+$0.33~dex.}
   \label{Fig.model}
         \end{figure*}

Second, in a 'classical' chemical evolution view of the Milky Way disc \citep[e.g.][]{tosi88, matteuccifrancois89, chiappini97, chiappini00, portinari98, boissierprantzos99}
 we do not expect that relatively young stars born in the inner disc to show an $\alpha$-enhancement. 
It is instructive to use a recent and simple chemical evolution model that does not include the 
effect of radial migration to study the most likely birthplace of the HAMR stars.
In Fig.~\ref{Fig.model} we show a  plot of  [Fe/H] and  [$\alpha$/Fe] vs. R$_{\rm GC}$  in the chemical evolution model of \citet{magrini09}. 
The three curves show the prediction for [Fe/H] and [$\alpha$/Fe] as a function of R$_{\rm GC}$ at present, 5~Gyr ago  and 10 Gyr ago.
The model predicts that the formation of a family of $\alpha$-enhanced metal-rich stars is very limited in space and in time. They formed in the inner part of the disc and during  the first epochs of disc formation (black curve), then 
the contribution of SNIa decreased the [$\alpha$/Fe] ratio to solar values. Thus, further stellar populations have higher [Fe/H] but lower [$\alpha$/Fe] than the oldest ones.
Similar results can be obtained even with more sophisticated models, as those by \citet{minchev13}, 
where the combination of high [$\alpha$/Fe]  and young ages is not predicted. For instance, in Figure~2 of \citet{minchev13}, stars with [O/Fe]=0.2 are older than 7 Gyr, 
whatever their birth location. Stars with slightly smaller [O/Fe] can be younger at the condition that they are born in the very outer disc, but then they would be metal poor.

We plot also [Fe/H] and  [$\alpha$/Fe] for the four clusters and for the solar neighbourhood and inner-disc samples in the same metallicity range of the clusters. We compute  [$\alpha$/Fe] considering [O/Fe], [Mg/Fe] and [Si/Fe]--the latter two corrected for the NLTE effects. 
For the field populations we assumed a distance of 8$\pm$1~kpc for the Solar neighbourhood stars, and of 5$\pm$2~kpc for the inner-disc sample. The uncertainties on the x-axes reflect their interval  of distances.

The four clusters are in good agreement with the model prediction in the [Fe/H] vs. R$_{\rm GC}$ plane, however in the [$\alpha$/Fe] vs. R$_{\rm GC}$ plane NGC~6705 is displaced with respect to 
what is expected for its age.  
Stars with the age of $\sim$0.3~Gyr, indeed, should have solar or sub-solar [$\alpha$/Fe] given their age and place of birth (blue curve of Fig.~\ref{Fig.model})
and thus their $\alpha$-enhancement is totally unexpected. Also the location of Be~81 is a bit surprising: as we discussed  above, this cluster does not have a high [$\alpha$/Fe] `per se'. 
However, if we consider its [$\alpha$/Fe] in conjunction with its [Fe/H], we find it a bit higher than expected. 
On the other hand, [$\alpha$/Fe]  in Tr~20 and  NGC~4815  are  consistent, within the errors, with the model curves (in between red and blue curve of  Fig.~\ref{Fig.model}). 

Therefore, the stars in NGC~4815 and Tr~20 have a [$\alpha$/Fe] that is fully consistent with their age and location in the disc as indicated by the predictions 
of the chemical evolution model of \citet{magrini09}. Be~81 has a slightly higher [$\alpha$/Fe] than expected for its [Fe/H] and R$_{\rm GC}$, 
However, the most striking case is that of NGC~6705,  which shows a genuine enhancement [$\alpha$/Fe] unlikely related to the HAMR stellar population in the solar neighbourhood, since they were born in a completely different epoch. 
Due to the young age of NGC~6705, its  $\alpha$-enhancement cannot be simply explained by disc chemical evolution, and also the hypothesis of \citet{magrini14} of radial migration from the inner disc is unlikely due to the young age of this cluster.  

Very recently, several studies revealed the presence of relatively young $\alpha$-enhanced stars.
\citet{chiappini15} report the discovery of a group of apparently young CoRoT red-giant stars exhibiting enhanced [$\alpha$/Fe] abundance ratios. These stars are more numerous in 
their inner-disc sample. Also \citet{martig15} analysing a  sample of giants  with seismic parameters from Kepler found a small percentage of stars with [$\alpha$/Fe]$>$0.2~dex and  ages below 4 Gyr. 
As in the case of our inner disc clusters, the existence of this kind of star cannot be explained in the framework of standard chemical evolution models of the Milky Way. 
Their presence might indicate that the chemical-enrichment history of the Galactic disc is more complex.
\citet{chiappini15} analyse some possible interpretations of the nature of these stars: {\em i)} they might have been formed close to the end of the Galactic bar, near co-rotation where gas stays inert 
for longer times and where the mass return from older inner-disc stellar generations is expected to be highest  with the net effect of  in-situ gas dilution; {\em ii)}  they might originate from a recent gas-accretion event. 
However, there is a substantial difference between the stars discussed in \citet{chiappini15} and the inner-disc open clusters: Chiappini's stars tend to be more metal poor than our clusters, with [Fe/H] ranging from $\sim-$0.6 to $\sim$0.1~dex, and they are on average older ($\sim$4~Gyr). 
Might the inner-disc open cluster be a younger tail of the "young' metal poor $\alpha$-enhanced population? 
Is the high [$\alpha$/Fe] of both populations related to their place of birth, near to co-rotation? 

In the next section, we try to estimate another possibility:  the effect of  a very local enrichment in the molecular cloud from which NGC~6705 was born to explain its chemical pattern. 

\subsection {The local enrichment hypothesis: the effect of a supernova explosion} 

Can a single SNII explosion explain the chemical pattern of NGC~6705 with its high [$\alpha$/Fe]? 
Using Gaia-ESO data, \citet{berg14} found a large scatter of [Fe/H] and  [Mg/Fe] of stars at each age  towards the inner disc. 
\citet{maline93} explained the observed scatter at each Galactocentric radius and age as due to the fact that the disc contains chemical inhomogeneities that can avoid remixing long 
enough. This allows the activation of the star formation process in both  enriched and unenriched portions, with typical timescales of 10$^{8}$-10$^{9}$ yr. 
\citet{gilmore89}  suggested that the scatter could be a consequence of self-enrichment in giant molecular cloud complexes. 
Their lifetime is short compared to the time scale of Galactic evolution, but much larger than the lifetime of the massive stars. 
Therefore, self-enrichment could occur inside a giant molecular cloud, and after the disruption of the cloud the metallicity 
inhomogeneity survives. It spreads only through a Galactic annulus, but it can be easily detected if the cluster formed from the molecular cloud 
is still observable. 

Starting from the assumption that the interstellar medium from which NGC~6705 was born had a mass comparable to its present time mass ($\sim$3500 M$_{\odot}$, Cantat-Gaudin et al. 2014), 
we compute the enrichment due to the explosion of a massive star of 15, 18 or 25~M$_{\odot}$ (yields from \citet{ww95}) in the hypothesis of a uniform
and instantaneous recycling of the medium. The enrichment has been computed
assuming Solar initial composition \citep{grevesse07}.

In Table~\ref{tab:enrichment}, we present the results of this calculation:  the explosion of a massive star with a mass 
15-18~M$_{\odot}$ might be able to explain the local enrichment of NGC~6705. 
The yields of \citet{ww95} can also explain the higher enrichment 
in [Si/Fe] with respect to the other elements. 
The hypothesis of local enrichment is supported also by 
\citet{roy95}  where characteristic timescales for homogenisation 
are short, 10$^6$-10$^8$ yr on scales of 100-1000 pc.
The number of  stars with a mass in the range 15-18 M$_{\odot}$ computed with the initial mass function of  \citet{cantat14} in the whole area of the cluster is about 2. 
Given the present total mass of NGC~6705, the expected number of stars with 15~M$_{\odot} <$M$<18~$M$_{\odot}$ 
is thus consistent with an early enrichment by one or more type II 
supernovae with such massive progenitors.  
However, the hypothesis that the SN was a cluster member
implies at least three assumptions; namely, {\it i)}
a prolonged star
formation (at least 10$^7$ Myr : timescales of SNII with progenitors of 15~M$_{\odot} <$M$<18~$M$_{\odot}$), 
where massive stars have formed first and have enriched the medium
from which the lower mass stars have subsequently formed; {\it ii)}
that the explosion of the SN has not dissipated the gas from which the low mass stars have formed; {\it iii)} that no low mass stars have formed from the
pristine (not yet enriched) gas, or that those stars have escaped the clusters. Similar problems are faced also in globular cluster formation and in their self-enrichment.

\begin{table}
\begin{center}
\caption{Cluster enrichment by SN II using yields of WW95.  }
\scriptsize 
\begin{tabular}{llll}
\hline\hline
Mass & [O/Fe] & [Mg/Fe] & [Si/Fe] \\
(M$_{\odot}$) & & & \\
\hline\hline
15 &     0.09   & 0.03 &     0.23\\
18  &    0.16  &  0.09  &    0.27\\
25   &   0.39 &     0.19  &   0.49 \\
\hline \hline
\end{tabular}
\label{tab:enrichment}
\end{center}
\end{table}

\section{Summary and conclusions}

We analyse the Gaia-ESO results ({\sc idr2/3}) for four inner disc clusters,
including the old cluster Be~81 which has not been studied in the Gaia-ESO
survey so far.
Using the improvements obtained in the metallicity measurement and the membership with RV determinations, we derive age, distance, and average reddening of Be~81 by using the classical approach of isochrone fitting.
The best-fit isochrone gives an age of $\sim$1~Gyr, a reddening $E(B-V)=$0.83 and a Galactocentric distance of 5.45~kpc.  
We compare the chemical pattern of Be~81 and of the other three old open clusters of {\sc idr2/3} with the solar-neighbourhood and inner-disc/bulge stars. 
We find that the  inner-disc/bulge giant stars and three open clusters show an enhancement in the {\em light} $\alpha$-elements (O, Si, Mg). 
We use the calibrator cluster M~67 to infer how much of this enhancement is related to NLTE effects, finding that the enhancement of [Mg/Fe] and [Si/Fe] in giants with respect to dwarfs is $\sim$0.10~dex. 
Applying these empirical NLTE corrections to  the chemical  patterns of open clusters, we find that Tr~20, NGC~4815, and Be~81 do not show any remarkable $\alpha$-enhancement. 
On the other hand, NGC~6705 still shows an uncommon pattern, with a significant $\alpha$-enhancement. 
We compare the cluster [$\alpha$/Fe] with those of the HAMR stars, finding that they belong to very different epochs of the lifetime of the Galaxy. 
We relate them also to the recently discovered population of young (age$\sim$4~Gyr) and $\alpha$-enhanced stars (\citep{martig15}, \citep{chiappini15}) mainly located towards the Galactic Centre. 
The inner-disc clusters, with ages between $\sim$0.3 to $\sim$1.5~Gyr, and the  "young" $\alpha$-enhanced stars might belong to an evolutionary sequence of the  same population, born in the inner part of the disc. The common chemical features of some of the clusters with the  "young" $\alpha$-enhanced stars might indicate peculiar conditions for the star formation near to the co-rotation. 
In our discussion, we also tentatively explain the very high [O/Fe], [Mg/Fe], and [Si/Fe] in NGC~6705 with an episode of local enrichment due to a type II supernova
in the mass range 15-18~M$_{\odot}$, as, e.g., contemplated by \citet{maline93} to explain the observed inhomogeneities at each Galactocentric radius.

\begin{acknowledgements}

The Authors thank an anonymous referee whose comments and suggestions improved the quality of the paper. 
The results presented here benefited from discussions in three Gaia-ESO workshops supported by the ESF (European Science Foundation) through the GREAT (Gaia Research for European Astronomy Training) Research Network Program (Science meetings 3855, 4127 and 4415).
This work was partially supported by the Gaia Research for European Astronomy Training (GREAT-ITN) Marie Curie network, funded through the European Union Seventh Framework Programme [FP7/2007-2013] under grant agreement n. 264895. 
We acknowledge the support from INAF and Ministero dell'Istruzione, dell'Universit\'a e della Ricerca (MIUR) in the form of the grant "Premiale VLT
2012" and The Chemical and Dynamical Evolution of the Milky Way and Local
Group Galaxies (prot. 2010LY5N2T).
V.A. acknowledges the support from the Funda\c{c}\~ao para a Ci\^encia e a Tecnologia, FCT (Portugal) in the form of 
the fellowship SFRH/BPD/70574/2010.
S.V. gratefully acknowledges the support provided       by Fondecyt. reg. 1130721.
T.B. was supported by the project grant ``The New Milky'' from the Knut and Alice Wallenberg foundation.
U.H. acknowledges support from the Swedish National Space Board (SNSB/Rymdstyrelsen).
G.T. acknowledges the support from the Research Council of Lithuania (grant No. MIP-082/2015).
This research has made use of the SIMBAD database, operated at CDS, Strasbourg, France.  
\end{acknowledgements}

\end{document}